\documentclass[11pt,onecolumn]{emulateapj}
\journalinfo{Ap.J. accepted Dec. 2005}

\newcommand{\et}{{\it et al.}}
\newfont{\myfont}{cmmib10}

\newcommand{\btheta}{\hbox{\myfont \symbol{18} }}

\shorttitle{Radio variability of Sgr A*}
\shortauthors{Macquart \& Bower}

\begin{document}

\title{Understanding the Radio Variability of Sgr A*}


\author{Jean-Pierre Macquart\altaffilmark{1}}
\affil{National Radio Astronomy Observatory, P.O. Box 0, Socorro NM 
87801, U.S.A. \email{jmacquar@nrao.edu}}
\author{Geoffrey C. Bower}
\affil{Astronomy Department and Radio Astronomy Laboratory, University of California, Berkeley, Berkeley, CA 94720, U.S.A.  \email{gbower@astron.berkeley.edu}}

\altaffiltext{1}{Jansky Fellow}

\begin{abstract}
We determine the characteristics of the 7\,mm to 20\,cm wavelength radio variability in Sgr A* on time scales from days to three decades.  The amplitude of the intensity modulation is between 30 and 39\% at all wavelengths.  Analysis of uniformly sampled data with proper accounting of the sampling errors associated with the lightcurves shows that Sgr A* exhibits no 57- or 106-day quasi-periodic oscillations, contrary to previous claims.  The cause of the variability is investigated by examining a number of plausible scintillation models, enabling those variations which could be attributed to interstellar scintillation to be isolated from those that must be intrinsic to the source.  Thin-screen scattering models do not account for the variability amplitude on most time scales.  However, models in which the scattering region is extended out to a radius of 50-500\,pc from the Galactic Center account well for the broad characteristics of the variability on $>4$-day time scales.  The $\sim 10$\% variability on $<4$-day time scales at $0.7-3\,$cm appears to be intrinsic to the source.
% Centimeter-wavelength variations on time scales less than $5\,$days appear to be intrinsic to the source.  
The degree of scintillation variability expected at millimeter wavelengths depends sensitively on the intrinsic source size; the variations, if due to scintillation, would require an intrinsic source size smaller than that expected. 
 \end{abstract}


%--------------------------------------------------------------------------------------------
\section{Introduction}
% INTRODUCTION
%--------------------------------------------------------------------------------------------
The compact radio source associated with the black hole at the Galactic Center, Sgr A*, is known to vary at millimeter and centimeter wavelengths on time scales from hours to years (Brown \& Lo 1982; 
%Genzel \et\ ???? ;
Zhao \et\ 1992; Bower \et\ 2002; Herrnstein \et\ 2004).  The origin of these variations remains unclear, with strong arguments for both extrinsic and intrinsic mechanisms having been advanced (e.g. Zhao \et\ 1989 compared with Zhao \et\ 2001).  

Interstellar scintillation is the primary mechanism which may cause any extrinsic variability.
The same plasma that is responsible for the scatter-broadening of Sgr A* at millimeter and centimeter wavelengths (e.g. Lo \et\ 1998, Bower \et\ 2004) is also expected to cause the source to exhibit refractive intensity variations.  It has been argued that much of the monthly to yearly variability in Sgr A* at wavelengths longer than 6\,cm can be explained in terms of refractive interstellar scintillation provided that scattering material moves across our line of sight at a relatively high speed of $\sim 1000-2000\,$km\,s$^{-1}$ (Zhao \et\ 1989).   However, the variability amplitude is not so easily accounted for: a scattering medium modeled as a single thin screen underpredicts the observed variability amplitude ({\it ibid.}) while extended medium models, which are in principle capable of explaining higher refractive modulation amplitudes for the same degree of scatter broadening, have not been investigated in the context of the Galactic Center.

Recent interpretations favor an intrinsic origin for much of the centimeter wavelength variability.  These center around claims of 106-day quasi-periodic variations at wavelengths shorter than $\sim 3\,$cm (Zhao \et\ 2001) and of 57-day quasi-periodic behavior at 2.3\,GHz (Falcke 1999).  The oscillations possess only a modest spectral purity, with the highest purity $\nu/\Delta \nu=2.2 \pm 0.3$ reported at 1.3\,cm.  Zhao \et\ (2001) discuss the origin of these oscillations in terms of periodic flares from a jet nozzle or an instability in the accretion disk triggering, for example, quasi-periodic production of convection bubbles.  It is widely supposed that the oscillations must reflect a process intrinsic to Sgr A* itself because scintillation is incapable of producing such regular oscillations.  Yet observations of certain intra-day variable quasars, whose variations are proven to be scintillation-induced,  invalidate this argument because their fluctuations often exhibit even higher degrees of spectral purity (Kedziora-Chudczer \et\ 1997; Rickett, Kedziora-Chudczer \& Jauncey 2002).  

Nonetheless there is little dispute that at least some of the variability is intrinsic.
Detections of flares at millimeter, IR and X-ray wavelengths (Wright \& Backer 1993; Tsuboi, Miyazaki \& Tsutsumi 1999, Eckart \et\ 2004; Baganoff \et\ 2001) conclusively demonstrate that the source is intrinsically variable.  A possible connection between X-ray flaring and unusually large flux density excursions at 7\,mm is also reported (Zhao \et\ 2004).  However, it is difficult to ascertain how much variability observed at centimeter wavelengths could be attributed to flaring since neither the duty cycle nor the energy distribution of mm or X-ray flares is well-constrained, much less the physical connection between centimeter and mm or X-ray behavior. 

%The literature is replete with pithy remarks about whether scintillation is a viable explanation for any of the variations observed in Sgr\,A*.  

Despite the many recent observational results concerning the properties of Sgr A*'s variability, a dearth of corresponding theoretical efforts has failed to place these results in context, leaving us none the wiser as to their cause.  For instance, while it is acknowledged that scintillation variability is likely to be important at centimeter wavelengths, no variations have been specifically attributed to it, and no realistic modeling has been applied to investigate what contribution it could conceivably make.  This paper aims to redress the balance by investigating two outstanding issues: (i) what exactly does a model of Sgr A*'s variability need to explain and (ii) can one deduce which variations {\it must} be intrinsic to the source by eliminating the variations that can be explained by interstellar scintillation?  The next section of this paper is devoted to the former question, including a critical examination of the $\sim 100\,$day quasi-periodic oscillations reported in Sgr A* (Zhao \et\ 2001), while \S\ref{Scint} addresses the latter question.  We compare the models to the observations in \S\ref{Comparison}, and summarize our findings and briefly detail their implications in \S\ref{Conclusions}.

%--------------------------------------------------------------------------------------------
\section{Data Analysis} \label{Data}
% DATA ANALYSIS
%--------------------------------------------------------------------------------------------
Sgr A* has been the subject of numerous VLA monitoring campaigns since 1975.  The resulting data are published in Zhao \et\ (1992, 2001) and Herrnstein \et\ (2004).  The latter lists the results of a three-year effort to measure weekly variations at 7\,mm, 1.3\,cm and 2\,cm.
We combine all these data to form lightcurves at 7\,mm, 1.3\,cm, 2\,cm, 3\,cm, 6\,cm and 20\,cm in an attempt to quantify the variability of Sgr A* on time scales of a few days to decades.  

Additional daily Green Bank Interferometer (GBI) monitoring at 2.3 and 8.3\,GHz (Falcke 1999) quantifies variations in Sgr A* on shorter time scales.  We reanalyze these data here, but do not incorporate them with the VLA flux density measurements because the GBI is highly susceptible to confusion in the Galactic Center region.  The GBI is a two-element interferometer whose 2400\,m spacing is insufficient to resolve out much of the extended emission near Sgr A* which, if not properly accounted for, can cause hour-angle dependent variations in the measured flux density of Sgr A*.   Falcke (1999) attempted to correct for hour-angle dependent gain variations, and to eliminate the contribution of confusion by comparing GBI flux density measurements with available contemporaneous VLA measurements.  But only by comparing the visibilities to a complete synthesis image of the crowded Galactic Center region can one be confident in removing the effect of confusion. 

\begin{figure}[h]
\begin{center}
\includegraphics[scale=0.9]{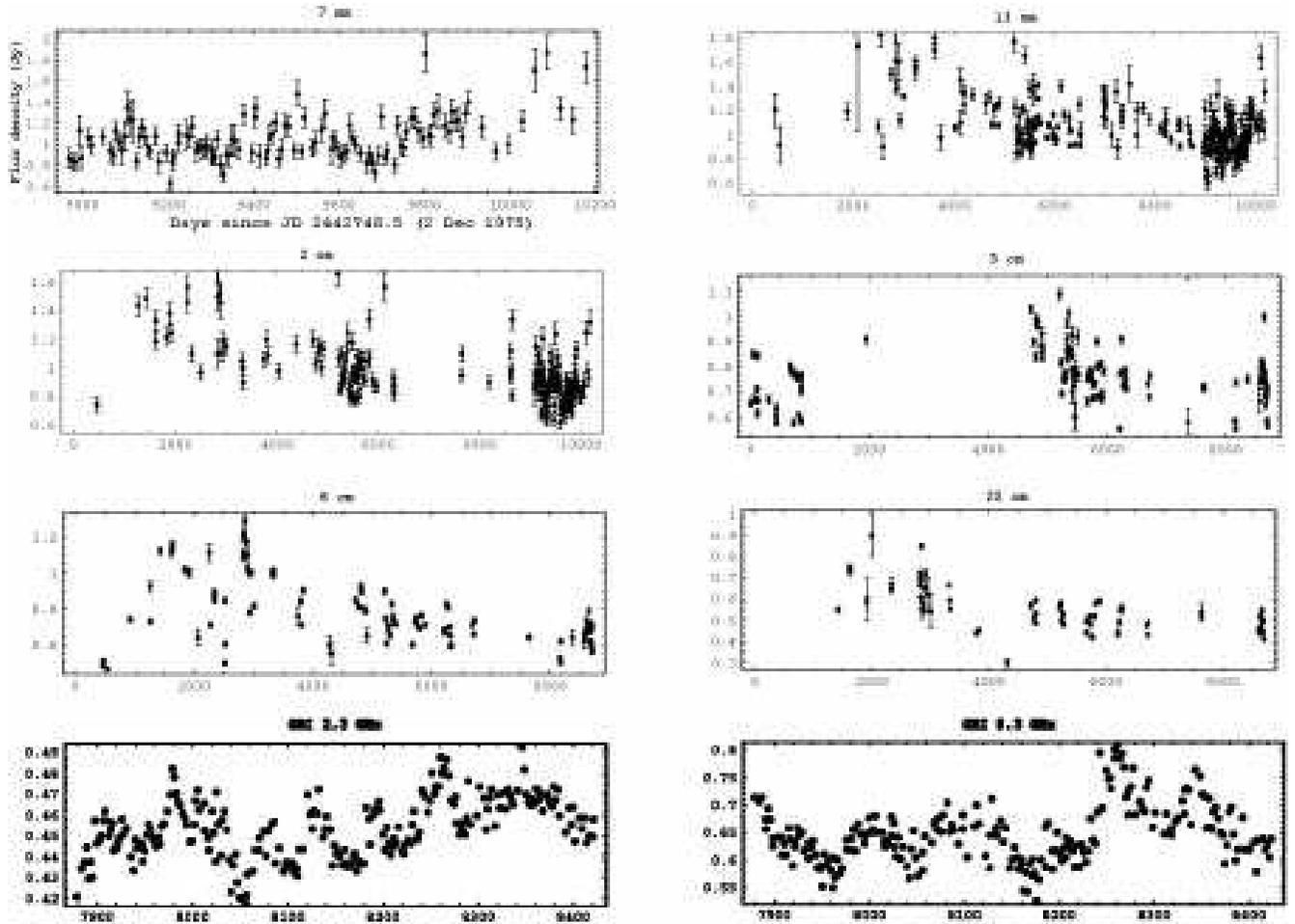} 
\end{center}
\caption{Lightcurves of the variations in Sgr A* from 7\,mm to 20\,cm.  Flux density errors were not included in the GBI data reduced by Falcke (1999), but the spread of the data points at each epoch is a reasonable indicator of the uncertainty. } \label{lightcurves}
\end{figure}

The lightcurves from the combined datasets are shown in Figure \ref{lightcurves}.  Various parts of the lightcurves have been published elsewhere, and their main purpose here is to illustrate exactly which data are used here.  

Throughout this paper we adopt the structure function as the chief measure of source variability because it is ideally suited to the interpretation of data which are highly irregularly sampled in time.  The intensity structure function, $D_I(\tau) = \langle [I(t'+\tau) - I(t')]^2 \rangle$, is a simple statistic which characterizes the variance between measurements separated by a time interval $\tau$.  Since we wish to be confident that we interpret only those features of the variability that are statistically significant, this requires a rigorous assessment of the errors associated with our measure of variability, and a statistic simple enough that the errors are readily calculable (see \S\ref{ErrorSection} below).  The power spectrum, which is related to the structure function by a Fourier transform, is a more elegant measure of variability, but we do not employ it here because the irregular time sampling of our datasets complicates the error analysis and thus the interpretation of the statistic. 

In computing a single structure function to characterize variability over the entire lightcurve we make the implicit assumption that the variability statistics are wide-sense stationary, which is to say that the statistical properties of the variations themselves do not vary with time.  This approach is not strictly valid, for instance, if the source undergoes various ``phases'' of variability in which the presence of fast time scale variations is modulated by some underlying long-term process;  X-ray binaries, whose behavior is characterized by infrequent outbursts, represent an obvious counterexample.  Although the possibility that variability in Sgr A* changes character with time cannot be discounted, there is no strong evidence to support the notion. 

% Ds saturates at some sufficiently long time scale.  
% so we make the assumption from the outset that this is not the case, and assume that whatever processes cause the variability always cause variability.

% discuss interpretation of peaks, dips, etc.?

\begin{figure}
\begin{center}
\includegraphics[angle=0,scale=0.7]{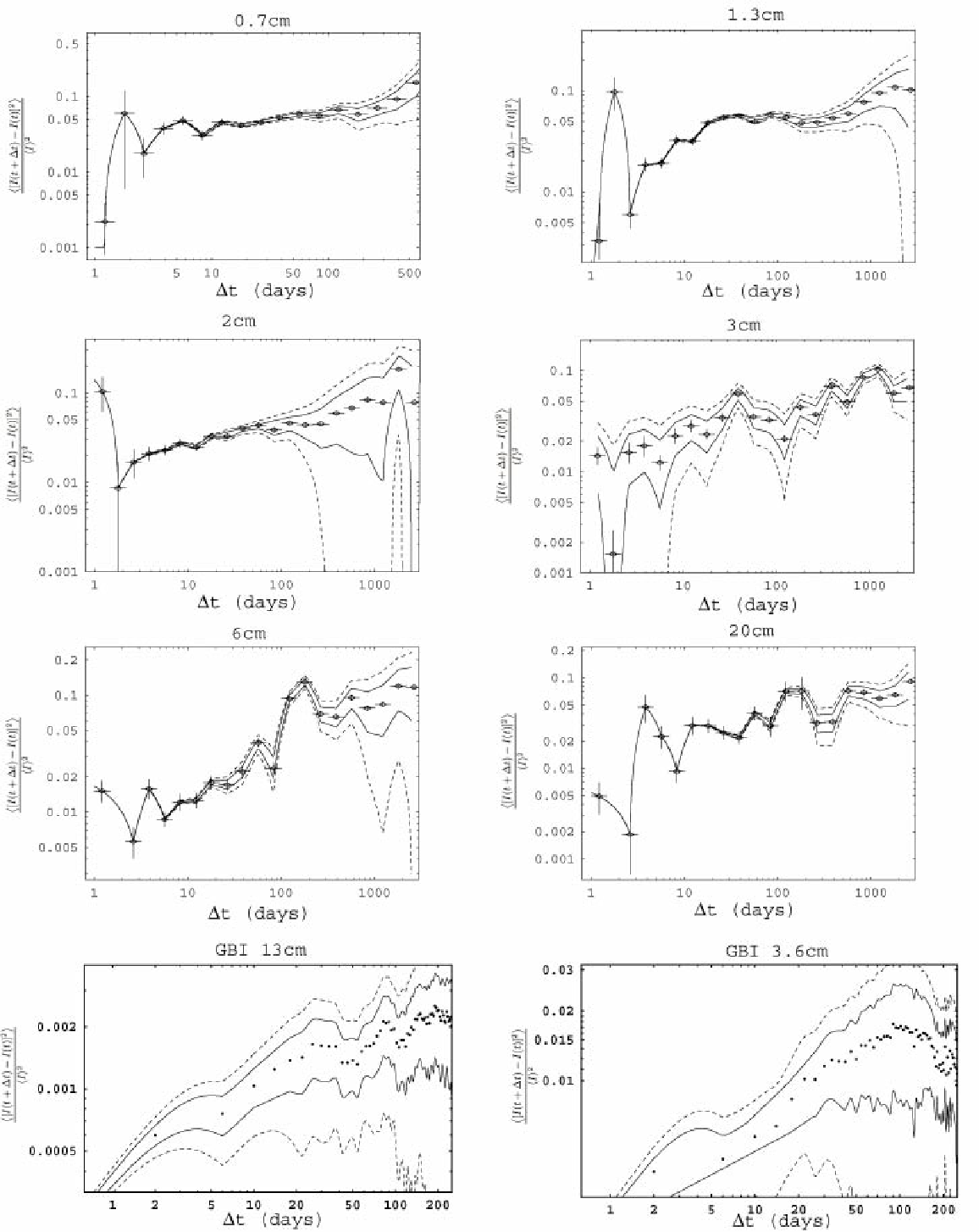} 
 \end{center}
\caption{Structure functions of the variations in Sgr A* from wavelengths of 7\,mm to 20\,cm.  The continuous and dashed error lines indicate the 1- and 2-$\sigma$ confidence limits of the measured structure functions due to the finite duration of the lightcurves (see \S\ref{ErrorSection}).  The ticks associated with individual points display the error caused by the finite number of observations that contributed to the measurement at that particular time lag.} \label{StructureFns}
\end{figure}

\begin{figure}
\centerline{
\includegraphics[angle=0,scale=0.6]{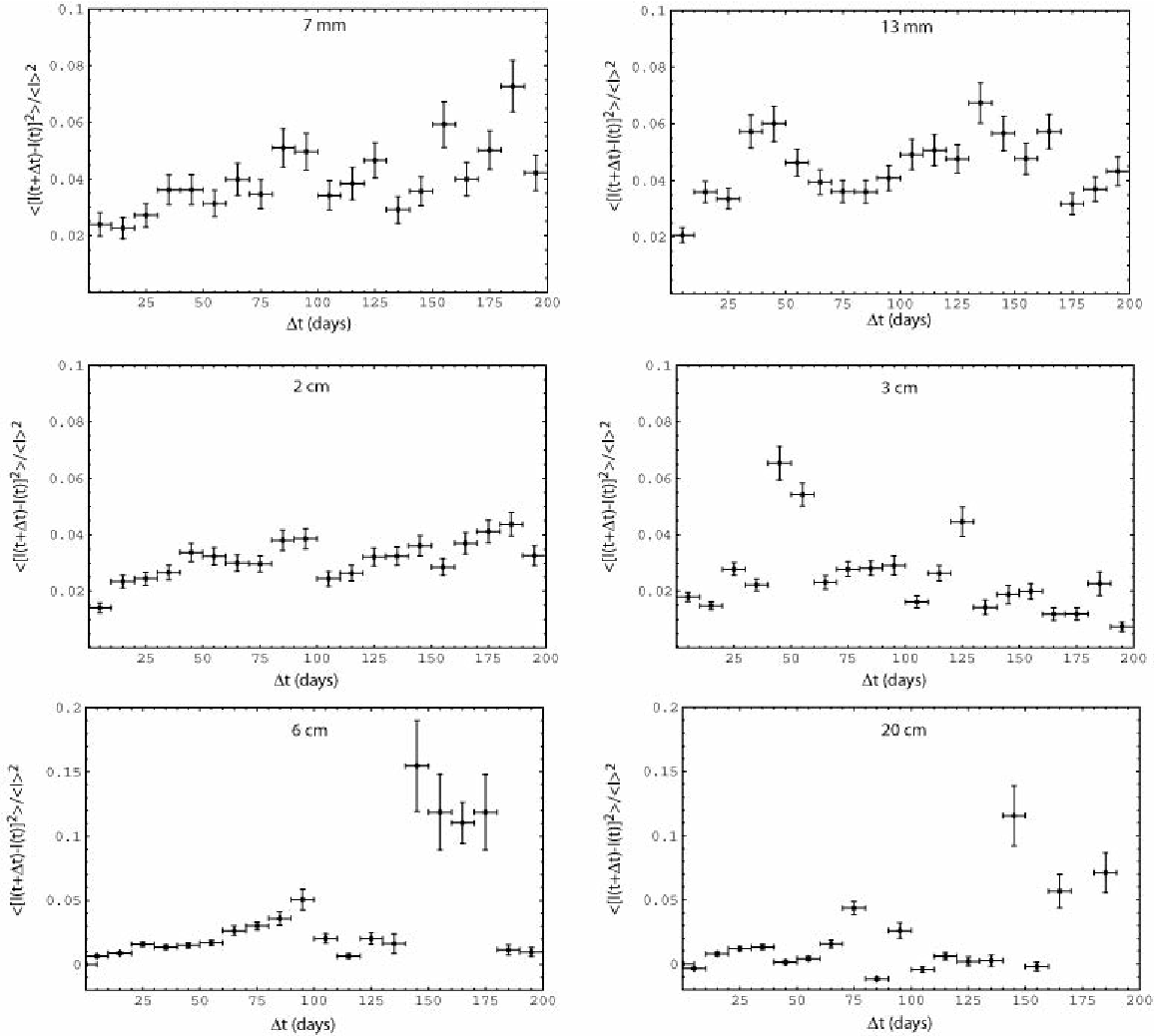}  
}
\caption{Structure functions of the variations in Sgr A* from wavelengths of 7mm to 20cm.  For small lags $\Delta t < 200\,$days the errors are dominated by the finite number of observations that contributed to the measurement at that particular time lag.  The contribution of measurement errors is estimated using the errors quoted from the observations, and has been subtracted from each structure function.  At 6cm and 20cm these clearly overestimate the true error, as these structure functions are negative at certain time lags.} \label{LinearDs}
\end{figure}

\begin{figure}
\centerline{
\includegraphics[angle=0,scale=0.8]{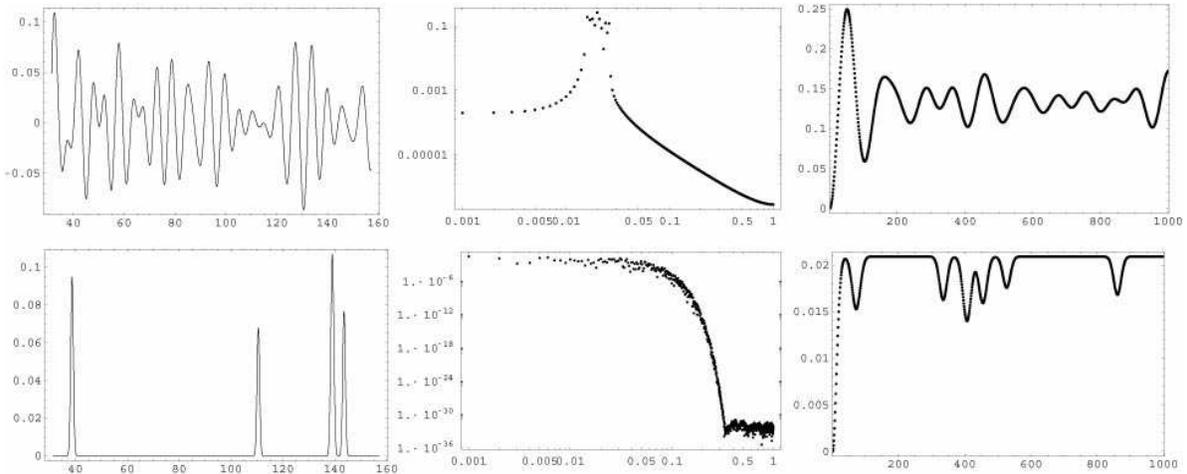} 
}
\caption{A simple illustration of the effect of quasi-periodic oscillations (top) and flaring (bottom) on the behavior of the structure function.  The left panels show the sample lightcurve, the middle panel its power spectrum and the right panels the resulting structure function.  The quasi-periodic oscillations possess a relative bandwidth $\Delta \nu/\nu$ of 0.5.} \label{QPOdiags}
\end{figure}

\subsection{Assignment of Errors} \label{ErrorSection}
% How we determined the errors assuming that they are drawn randomly 
% form a wide-sense stationary ensemble.
%--------------------------------------------------------------------------------------------
The correct determination of errors associated with any measure of the variability is crucial in assessing the significance of variability time scales or even periodicities in the data, particularly when they are irregularly sampled and when  
the time scales under consideration are comparable to the entire data length.  It is also crucial when comparing the observed variations to theoretical models, as most theories only predict ensemble-average quantities (i.e. they predict the climate, not the weather). 

The largest contribution to the error arises because our observations only sample the stochastic fluctuations over a finite duration.  A simple argument would suggest, for instance, that in a dataset spanning 1000 days a process that operates on a time scale of 200\,days  contains {\it at most} five independent measurements of the variations caused by this process.   Formal arguments show that the error in the structure function at delay $\tau$ from an observation of total duration $T$ is (Jenkins and Watts 1968, see also Appendix B of Rickett, Coles \& Markannen 2000)
\begin{eqnarray}
var[D(\tau)] = \frac{4}{(T-|\tau|)^2} \int_{0}^{T-\tau} \left[ \gamma (|r-\tau|)  +\gamma(|r+\tau|) - 2 \gamma(r) \right]^2 (T-\tau-r) dr,
\end{eqnarray}
where the function $\gamma(\tau)$ is the ensemble-average autocovariance of the intensity fluctuations at wavelength $\lambda$.  This function is unknown, so we assume that the measured autocovariance is a reasonable representation of its corresponding ensemble-average counterpart.  
%  NB the result is a factor of 4 lower if we use structure functions instead of auto-correlations in the above equn.

%Similarly, Bartlett's formula (Jenkins \& Watts 1968) characterizes the error in the cross-correlation between two datasets, here labeled by suffixes 1 and 2,
%\begin{eqnarray}
%var[C_{12}(\tau)] = \frac{T-\vert \tau \vert}{T^2} \int_{-T+|\tau|}^{T-|\tau|} \left[ \gamma_{11}(r) \gamma_{22} (r) + \gamma_{12}(r+\tau) \gamma_{12}(-r-\tau) \right] \left(1 - \frac{|r|}{T-|\tau|} \right) dr,
%\end{eqnarray}
%where $T$ is the span of the datasets, assumed identical here, and $\gamma_{ij}(\tau),$ is the ensemble-average auto- or cross- correlation between datasets $i$ and $j$.  Again, since the ensemble average auto- and cross-correlations are unknown we bootstrap by assuming that the measured auto- and cross-correlations are reasonable representations of their corresponding ensemble-average counterparts.

%%The issue is further complicated if additional variability occurs on longer time scales.  The long time scale process may be statistically independent when averaged over the ensemble of all possible variations, but chance correlations on a sufficiently short dataset can cause power to leak between these two nominally independent oscillations.

%--------------------------------------------------------------------------------------------
\subsection{Characteristics of the variability} \label{VarCharacteristics}
% Lack of quasi-periodic oscillations.  
%---------------------------------------------------------------------------------------------
The structure functions derived from the lightcurves are displayed in Figures \ref{StructureFns} and \ref{LinearDs}.  Figure \ref{StructureFns} is the main result of this section, but the plots in Figure \ref{LinearDs}, shown on linear scales, allow closer scrutiny of the variability characteristics on time scales shorter than 200 days. 
  
\subsubsection{Intra- to inter-day fluctuations} \label{IntraDay}

Sgr A* is reported to exhibit variations on scales down to less than one day (Brown \& Lo 1982).  It is possible to determine whether the present datasets show evidence for intra-day ($\leq 4$-day) flickering by examining the behavior of the structure functions at small time lags.  Short time scale flickering is present if the value of the structure function in the smallest time bin differs significantly from zero.  A proper assessment of the presence of flickering depends crucially on the correct determination of the errors associated with the flux density measurements.  We have subtracted the contribution of measurement errors from each structure function using the errors quoted in the papers from which the observations are derived\footnote{The means of removing measurement errors is obvious when one writes the structure function in terms of the autocovariance: $D(\tau) = 2 [C(0) - C(\tau)],$ where $C(\tau) = \langle \Delta I(t+\tau) \Delta I(t) \rangle$.  Measurement noise is assumed to be uncorrelated between samples and independent of the true intensity fluctuations, so it only makes an (additive) contribution to $C(0)$.  It is removed by subtracting the variance of the measurement errors from $C(0)$.}.  However, if the errors are under-estimated the structure function is biased towards high values, leading to an over-estimate of the amplitude of short time scale flickering.   Conversely, an over-estimate of the errors biases the structure function to lower values, the most clear manifestation of which is to cause dips in the structure function below zero, which is clearly unphysical.

The structure functions in Fig.\,\ref{LinearDs} suggest that Sgr A* undergoes appreciable flickering on  $<10$-day time scales at wavelengths from 7\,mm to 3\,cm.  In all cases the root-mean-square fluctuations are approximately $10$\% of the mean flux density.  Inspection of more finely-binned structure functions reveals that Sgr A* exhibits 10\%, 6\% and 8\% variability at 7, 13 and 20\,mm respectively on $<4$-day time scales.   However, we caution that our estimate of the flickering amplitude depends critically on a correct assessment of the errors associated with the observations.  Such an estimate is called into question for the 20\,cm lightcurves, for instance, where the estimated contribution of measurement errors leads to negative structure function values at certain lags.

The amplitude of the intra-day variability must be subtracted when comparing the observed structure functions to models which only apply to variations on longer time scales.  However, the amplitude of any intra-day variation is small compared to total amplitude of the intensity variations, and this correction is small compared to the uncertainty in the total variability amplitude on long time scales at most wavelengths.

\subsubsection{Quasi-periodic variations}

Figure \ref{LinearDs} can also be used to assess whether any variations exhibited by Sgr A* are quasi-periodic.  Such variability is characterized by the presence of oscillations in the structure function.  The origin of this behavior is understood by noting that the structure function is related to the power spectrum of the lightcurve, $P(\omega)$, by a Fourier transform: $D_I (\tau) \propto \int [1 - \exp(i \omega \tau)] P(\omega) d\omega$.  A purely sinusoidal signal in the lightcurve would manifest itself as a sharp peak in the power spectrum, and the structure function would exhibit a peak at a time scale corresponding to the period, followed by sinusoidal oscillations peaking at multiples of the fundamental period.  The amplitude of these oscillations would decrease with time lag if the variations are spectrally impure. 

It is important to distinguish between oscillations in the structure function from spikes which are devoid of accompanying oscillations at longer time scales.  A structure function containing sharp, isolated spikes indicates that the {\it power spectrum} contains quasi-periodic features.  This in turn indicates the presence of sharp spikes in the corresponding lightcurve.  The structure function contains spikes (or sharp dips) when at least two flares are present in the lightcurve.  For instance, flares at times $t_1$ and $t_2$ each of duration $\Delta t$ would give rise to a feature in the structure function at a time lag $|t_2-t_1|$ of width $2 \Delta t$.  In the absence of any quasi-periodic or flaring behavior the structure function is expected to increase monotonically with time until it saturates at the longest time scale of the variations present in the data.  Figure \ref{QPOdiags} illustrates how quasi-periodic oscillations and flares are manifested in structure functions.

We investigate the form of the structure function on $<150\,$day time scales, motivated by reports of quasi-periodic variations on time scales of $\sim 106\,$days between 7\,mm and 3\,cm (Zhao \et\ 2001).  The structure functions derived from the long duration VLA datasets (Fig. \ref{LinearDs}) are most useful in assessing the data for the presence of any unusual features.   A simple test for the significance of any features is obtained by fitting a single line through each structure function at time lags from $\Delta t=5-150\,$days and computing the reduced $\chi^2$ statistic, as listed in Table 1.  The departure of the structure function from a line, indicated by a high $\chi^2$, signifies the presence of peaked features above the generally-increasing trend with time.  The statistics in Table 1 show that the $7\,$mm structure function is well-fit by a single line, and there is no significant detection of any quasi-periodic variability. However, the reduced $\chi^2$ statistic suggests the presence of significant deviations at all lower wavelengths, as is obvious by inspection of Fig. \ref{LinearDs}.  The 1.3, 2 and 3\,cm structure functions appear to exhibit peaks at lags of $\Delta t=30-45$, $45-65$ and $45-55$ days respectively.  The 2\,cm structure function also exhibits a peak at $\Delta t=80-95\,$day which appears to coincide with a marginally significant peak at a corresponding time lag in the 7\,mm structure function, however the coincidence does not necessarily increase the significance of the peaks because the errors bars at the two wavelengths are not independent if the two lightcurves are partially correlated.  In addition, the 6 and 20\,cm structure functions possess highly significant peaks at $\Delta t \approx 140-170 $\,days; the 1.3\,cm structure function appears to exhibit a feature at similar time lags, but inspection of the error bars in Fig. \ref{StructureFns} suggests the detection is of marginal significance.

We also consider the lightcurves measured with the GBI.  Our reanalysis of the GBI dataset reproduces the structure functions reported by Falcke (1999) on whose basis a 57-day quasi-periodic variability cycle is claimed at 2.3\,GHz.  These variations were suggested to be quasi-periodic because the structure function subsequently oscillates weakly after peaking at $\sim 57$\,days.   However, our estimate of the errors associated with these structure functions casts doubt on the significance of any claim of quasi-periodic behavior.   At both 3.6 and 13\,cm even the 1-$\sigma$ error troughs are consistent with structure functions that increase monotonically and subsequently saturate without undergoing any oscillatory behavior.  The insignificance of the quasi-periodic behavior is not affected by changes in the temporal binning of the structure function.   The error troughs reflect only the error incurred  by trying to infer the ensemble average behavior of the variations from a dataset of finite duration.

\begin{figure}[h]
\includegraphics[angle=0,scale=0.75]{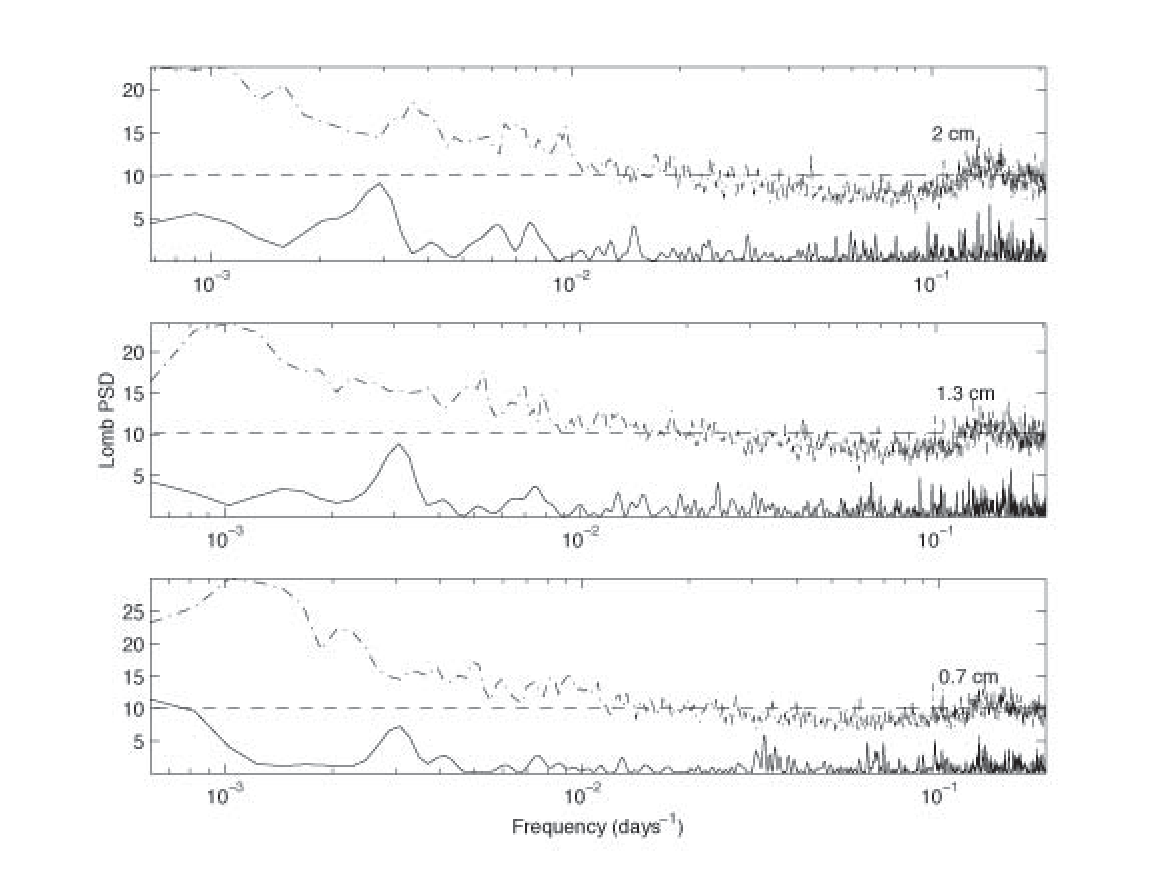}
\caption{Lomb periodograms of the variations between 2000 and 2003 at 20, 13 and 7\,mm.
The power spectral density for the data (solid line) is plotted with the 99th percentile expectation of uniform noise (dot-dashed line) and noise with a red spectrum (dashed line).  
} \label{LombFig}
\end{figure}

In summary, all of the particular features identified in the VLA structure functions are single, isolated peaks.  None of these can be attributed to oscillatory behavior. As a further, more sensitive test for quasi-periodic variability we present in Fig.\,\ref{LombFig} Lomb periodograms for 2 cm, 1.3 cm and 0.7 cm lightcurves based on well-sampled data from 2000 to 2003.  We compare the power spectral density for the data against the 99th percentile expectation of uniform noise (dot-dashed curve) and noise with a red spectrum (dashed line).  The red spectrum is calculated using 300 Monte Carlo
simulations that use the sampling function of the data sets.  We see clearly that there are no significant periods in the data.  In particular, there are no peaks in the vicinity of the 106-day period reported by Zhao \et\ (2001).  The spikes in the 2\,cm and 1.3\,cm data at $\sim 0.003 {\rm\ days}^{-1}$ have a significance of only $\sim 90$\% for the uniform noise case.  Moreover, these periods have only been sampled $\sim 3$ times in this data set and they are not apparent in PSDs from longer (but less well-sampled) data sets.  Against the red noise case, these peaks have minimal significance.  Results are similar to the uniform noise case when one estimates the PSD significance through Monte Carlo simulations in which the lightcurves are generated through reordering of the data. We conclude, then, that there is no evidence for periodic or quasi-periodic oscillations in the radio light curves of Sgr A*.

\subsubsection{Essential characteristics of the variability}

The structure functions presented in Fig. \ref{StructureFns} are the key observable that any variability theory must reproduce.  Broadly speaking, each structure function may be characterized by a monotonically increasing portion until it saturates at a  certain amplitude and time scale.  A viable explanation of the variability should explain the shape of the structure function over an appreciable range of time lags.  Even if the model does not explain every significant peak and wobble in the structure function, it is still viable if it explains the (i) amplitude and (ii) time scale at which the structure function saturates and (iii) the slope of the structure function over an appreciable range in time lags.  In the following sections we gauge the success of our model by its ability to reproduce these three characteristics.

As remarked above, many of the structure functions exhibit spikes. These are caused by large, rapid and isolated flux density excursions in the lightcurves.  We shall not attempt to explain these features in this paper.  We merely remark that they likely represent either flaring activity intrinsic to Sgr A* itself, or a manifestation of Extreme Scattering Events (Fiedler \et\ 1987) which appear to occur when the Earth traverses a caustic surface of certain lens-like discrete objects in the interstellar medium (Romani \et\ 1987; Walker \& Wardle 1998).
The prevalence of these flares in most of the structure functions indicates that the mechanism responsible operates over a factor of ten range in wavelength.  

%We shall see below that these features cannot be reproduced by scintillation models considered below, and must be produced by a qualitatively different phenomenon.  

\begin{table}
\begin{tabular}{|c|c|c|c|c|}
\tableline
$\lambda$ & saturation & saturation  & index of power law increase & reduced $\chi^2$ from linear fit \\
(cm) & amplitude & time scale (days) & before saturation &  to $D_I$ between 5 \& 150 days \\
% & reduced $\chi^2$ (fourth-order polynomial) \\
%& probability of getting $\chi^2$ as high as this by chance \\
\tableline
0.7 & 0.05 & 6 & $1.2 \pm 0.5$ & 1.92 \\ 
% & 1.36 & 0.17\\
1.3 & 0.05 & 30 & $0.7 \pm 0.2$ & 4.33 \\
%& 2.25  & 0.008  \\
2 & 0.05 & 50 & $0.8 \pm 0.15 $ & 4.2 \\
% & 1.87  & 0.03 \\
3 & 0.1 & 40 & $1.0 \pm 0.2$ &15.6  \\
%& 14.6 & 0 \\
6 & 0.1 & 200 & $1.0 \pm 0.2$ & 12.5 \\
% & 8.1 & 0\\
20 & 0.1 & 100-1000? & $1.6 \pm 0.6$ & 34.4 \\ 
% &  22.6 & 0 \\
\tableline
\end{tabular}
% nb last column computed from Chi-sq distn with 14 d.o.f. e.g.
% f[x_]= PDF[ChiSquareDistribution[14],x] 
% prob = 1- 14* Integrate[PDF[14 x],{x,0,1.2}]
\caption{Characteristics of the structure functions in Figures \ref{StructureFns} \& \ref{LinearDs}.  The index of the power law in the increasing part of the structure function is obtained from a fit of the function $A+ B \tau^\alpha$ to the data between days 5 and 40 in the logarithmically-binned structure functions.  The constants $B$ and $\alpha$ are allowed to vary, but the offset $A$ is chosen to remove any contribution that reflects intra-day variations whose contribution to the structure function does not appear to be associated with the variations on time scales longer than 5\,days (i.e. intra-day variability); this parameter is nonzero at 7, 13 and 20\,mm only (see the discussion in \S \ref{IntraDay}). 
The last column lists reduced $\chi^2$ values resulting from attempts to fit each structure function at lags between $5$ and $150$\,days by a straight line.}
\end{table}

%--------------------------------------------------------------------------------------------
\section{Scintillation Variability} \label{Scint}
% Use a refractive scintillation model and see what it explains so we know
% what must be attributed to intrinsic variability
%
% How much of these variations could be due to scintillation? 
%--------------------------------------------------------------------------------------------

In this section we attempt to distinguish between variations intrinsic to Sgr A* itself and those due to refractive interstellar scintillation.  Several detailed scintillation models that span the range of possible scattering conditions are constructed in order to isolate those variations which cannot be explained under any plausible scattering conditions, and must therefore be attributed to intrinsic source variability. The distinction between intrinsic and extrinsic variability is made on the basis of a scintillation model because the physics of any centimeter-wavelength intrinsic variability in Sgr A* is ill-constrained; indeed even the fraction of the radio emission originating in the jet and accretion disk is disputed (e.g. Falcke \& Markoff 2000; Quataert \& Narayan 1999; Yuan, Markoff, Falcke 2002).  On the other hand, the basic physics of interstellar scintillation is well-understood and makes robust predictions that can be compared directly to the observed intensity variations.  

The shortcoming of this approach lies in the uncertainty of the exact distribution of scattering material along the line of sight (Lazio \& Cordes 1998; Yusef-Zadeh \et\ 1994), which in turn affects the amplitude and time scale of the predicted variations.  To encompass the range of variations possible we consider a model in which the scattering material is entirely located in a single thin screen, either at a distance of $50$ or $500$\,pc from Sgr A*, and one in which the material is distributed in an extended medium near Sgr A*, again with a scale length of either $\Delta z =50$ or $500\,$pc.  
% more comments on extremely strong scattering in GC region?

Only variations caused by refractive interstellar scintillation are investigated here.  Fluctuations caused by diffractive scintillation are possible in principle, but the extremely strong scattering observed toward Sgr A* renders the time scale of such scintillation of order seconds at centimeter wavelengths.  Such variability is expected to be strongly quenched, given that recent estimates suggest the intrinsic source size far exceeds the angular scales probed by this phenomenon (Bower \et\ 2004). 

%(Caveat: their measurement of an overall source size does not preclude the existence of smaller internal substructure.)

%--------------------------------------------------------------------------------------------
\subsection{The scattering model}
% Introduce the extended medium and thin screen models we will use
%--------------------------------------------------------------------------------------------
The distribution of scattering material along the line of sight to Sgr A* is described in terms of the power spectrum of electron density fluctuations.  This is modeled in the following standard form:
\begin{eqnarray}
\Phi_{N_e} ({\bf q},z) = C_N^2 (z)\, \left( \frac{q_x^2}{R} + R q_y^2 \right) ^{-\beta/2} \exp \left[ - \left( \frac{q l_0}{2} \right)^2 \right], \label{PhiNe}
\end{eqnarray}
where the amplitude of the power spectrum is written as a function of the spatial wavenumber, ${\bf q}$, and the distance, $z$, from the source.  The quantity $l_0$ is the inner scale of the fluctuation spectrum, and is usually identified with the turbulent dissipation scale.
  
The $\zeta \approx 2$ axial ratio observed in the scatter-broadened image of Sgr A* (Lo \et\ 1998) indicates that the amplitude of the power spectrum varies with direction on the sky, presumably reflecting the orientation of the local magnetic field.  The observed anisotropy is the result of either a change of the inner scale or  the amplitude of the power spectrum as a function of orientation on the sky.  We concentrate on the latter case because such anisotropy is expected for MHD turbulence (Goldreich \& Sridhar 1995). 
In eq. (\ref{PhiNe}) the anisotropy is characterized by means of the parameter $R$, and is oriented so that the major axis of the scattering disk is along the $x$-axis.  The anisotropy of a scattered image, $\zeta$, is  equal to the anisotropy parameter $R$ when the length scales probed by angular broadening are much larger than the turbulent dissipation scale.  On smaller scales the two measures are approximately equivalent (see \S\ref{CN2determine} below). 
% When the diffractive scale is much larger than the inner scale, the ratio of $r_{\rm diff}$ measured along the $y$-axis to that measured along the $x$-axis is exactly $R$. 
%
% NB it would seem that $C_N^2$ is greater along the y-axis and thus that the scattering disk is along the
% y-axis.  But numerical integrations show it's the other way around.  To see this, just write the spectrum:
% propto 1/ (q_x^2/R + R q_y^2)^\beta.  So for q_y =0 we get an R on the numerator associated with q_x.

% The case in which only the inner scale varies is uninteresting because refractive scintillation is sensitive to the low-wavenumber end of the turbulence power spectrum, where the inner scale is irrelevant. In this case the scintillations would resemble those from isotropic turbulence.   

The spectral index of the turbulence is assumed to be independent of $z$, but its amplitude is allowed to vary through the quantity $C_N^2(z)$.  For a thin screen located at a distance $z_0$ from Sgr A* one writes,
\begin{eqnarray}
C_N^2 (z) = \left\{ \begin{array}{ll} 
C_N^2, &   |z-z_0|  < \Delta L/2\\
0, & \hbox{ otherwise}, \end{array} \right.
\end{eqnarray}
where the thickness of the medium, $\Delta L$, is assumed to be far less than the source-observer distance, $L$.  In the thin screen model the scattering measure ${\rm SM}=C_N^2 \Delta L$ and the screen distance are the free parameters.  

When the scattering measure is large, as it is in the Galactic Center environment (see \S\ref{CN2determine} below), the refractive modulations from a thin screen can be small.
To this end, we also investigate a model in which the turbulent fluctuations are extended along the line of sight.  Extended media are capable of producing larger refractive modulations relative to thin screens (Romani, Narayan \& Blandford 1986; Coles \et\ 1987).  We consider the specific distribution in which the amplitude of the turbulence declines as a Gaussian from the Galactic Center:
\begin{eqnarray}
C_N^2 (z) = C_N^2 (0) \exp \left( - \frac{z^2}{\Delta z^2} \right). 
\end{eqnarray}
The free parameters of this model are the scale length of the distribution, $\Delta z$, and a normalization constant that sets the overall amplitude of the turbulence, $C_N^2(0)$.  The effective scattering measure of the medium in this model is  ${\rm SM}=C_N^2(0) \pi^{1/2} \Delta z /2$ when $\Delta z$ is much smaller than the distance to the source.

Unfortunately, the additional complexity inherent to the treatment of fluctuations in an extended medium forces us to abandon the explicit inclusion of anisotropy for this model.  We take $R=1$ and normalize the model to a scattering strength intermediate to that implied by the scatter-broadening along the major and minor axis of Sgr A*.  We feel that our failure to take anisotropy into account in a self-consistent manner in this model is a minor point compared to other uncertainties, such as the true distribution and strength of scattering material along the line of sight.  In any case, the correct manner in which to incorporate anisotropy in an extended scattering medium model is highly uncertain.  Goldreich \& Sridhar (1995) point out that that the local value of $R$ in a thin plane of MHD turbulence, where ``thin'' is less than the outer scale of magnetic field fluctuations, is expected to be $\ga 10^3$, and the value of anisotropic image broadening that one actually observes is much lower only because it represents an average over many different orientations of the magnetic field along the line of sight\footnote{The Goldreich \& Sridhar (1995) theory of Kolmogorov MHD turbulence suggests that the local anisotropic ratio is $\sim (k_\perp L_{\rm out})^{1/3}$, where $L_{\rm out}$ is the outer scale of the turbulent magnetic field fluctuations and $k_\perp$ is the wavenumber on which the power spectrum is probed.}.

% see section 6.2 of Goldreich and Sridhar 1995.

\subsection{Determination of $C_N^2$ appropriate to the models} \label{CN2determine}

The angular size of the scatter-broadened image of Sgr A* provides a direct means to determine the scattering strength appropriate for our scattering models.  For a source of unit intensity whose intrinsic angular size is much smaller than the angular broadening size, the visibility\footnote{We denote this observed visibility by $\Gamma$ to distinguish it from the intrinsic source visibility, $V$ used below.} observed on a baseline ${\bf s}$ is
\begin{eqnarray}
\Gamma({\bf s}) = \exp \left[ -\frac{1}{2} \int_0^L D_\phi'({\bf s} z/L,z) \, dz \right]. \label{MutCoher}
\end{eqnarray}
The quantity $D_\phi'$ is the derivative of the phase structure function with respect to $z$ and is given by
\begin{eqnarray}
D_\phi'({\bf s},z) = 4 \pi r_e^2 \lambda^2 \int_{-\infty}^{\infty} \Phi_{N_e}({\bf q},z)\, (1-e^{i {\bf q} \cdot {\bf s}} ) \, d^2{\bf q}. \label{DphiPrime}
\end{eqnarray} 
The baseline ${\bf s}_0$ at which the visibility falls to $1/e$ of its maximum value at wavenumber $k=2 \pi/\lambda$ is related by $\btheta_0=2/k {\bf s_0}$ to the angular radius at which the brightness distribution falls to $1/e$ of its maximum value, $\theta_0$.
% \footnote{This is easily seen when the visibility has the form $V(s) = \exp(-s^2/2 a^2)$.  The $1/e$ point is at $s_0=\sqrt{2} a$. 
%The image brightness is then $I(\theta) \propto \exp(- k^2 \theta^2 a^2/2)$.  So the $1/e$ point of the brightness distribution occurs at $\theta = \sqrt{2} /k a =  \sqrt{2}/(k s_0/\sqrt{2}) = 2/k s_0$.}.  
One therefore solves $\int_0^L D_\phi'(2z/k \btheta_0 L,z)dz=2$ to determine the scattering strength.
% NB 1/e point of visibility at s_0 is related to \theta_0  by 
%\theta= 1/k s_0 (where the brightness of the scattered image falls to $e^{-1}$ of its maximum value).  
% It's easy to see this: the visibility goes as V(s) = exp[-s^2/2 a^2].  1/e point is at s_0=sqrt{2} a. 
% FTing this gives I(theta) propto exp[- k^2 \theta^2 a^2/2].  So the 1/e point of the brightness distn is
% \theta = \sqrt{2} /k a = theta_0 = \sqrt{2}/(k s_0/sqrt{2}) = 2/k s_0.

%The scattering models are normalized with the condition
%\begin{eqnarray}
%2= \int_0^L D_\phi'({\bf s}_0 z/ L,z) \, dz = 4 \pi r_e^2 \lambda^2 \int_0^L dz \int  \Phi_{N_e}({\bf q},z)\, (1-e^{i z\,{\bf q} \cdot {\bf s}_0/L}) \, d^2{\bf q} ,
%\end{eqnarray}
% NB \Gamma(s) = 1/e = exp[-D(s)/2].  So this requires us to solve D(s) = 2 
% which is equivalent to D(1/k\theta) = 2.

Measurements of angular broadening probe the structure function on a length scale $\sim s z/L=2 z/k \theta_0 L$, which is of order kilometers for the case of Sgr A*.  Consider, for instance, the scale probed by angular-broadening at 6\,cm, where the angular diameter of Sgr A* is $49.6\,$mas along its major axis.  Assuming the scattering occurs at $z=100\,$pc and taking the distance to the Galactic center as $L=8.5\,$kpc, one finds that angular broadening is sensitive to structure on scales of only $2 \times 10^3\,$m.  

This scale is much smaller than the expected inner dissipation scale.  Spangler \& Gwinn (1990) argue that the inner scale of the turbulent cascade is plausibly identified with the larger of the ion inertial length 
\begin{eqnarray}
l_i= \frac{v_A}{\Omega_i} = 228 \left( \frac{\rho}{1\,{\rm cm}^{-3}}\right)^{-1/2} \, {\rm km},
\end{eqnarray}
or the ion Larmor radius 
\begin{eqnarray}
r_i = \frac{v_{\rm th}}{\Omega_i}= 300 \left( \frac{T_i}{10^3\,{\rm K}} \right)^{1/2} \left( \frac{B}{1\,\mu{\rm G}} \right)^{-1} \, {\rm km},
\end{eqnarray}
where $T_i$ is the ion temperature.  The dissipation scale is larger than several kilometers for the range of plausible densities, temperatures and magnetic fields in the Galactic Center.

The length scale on which angular broadening probes the scattering medium is important in determining the scattering strength because the character of the phase structure function changes when its argument falls below the turbulent dissipation scale.  The structure function scales as $s^{\beta-2}$ above this point and as $s^2$ below it.  Defining $s_{\rm diff}$ as the length scale over which the rms phase change is one radian, one sees that $s_{\rm diff} \propto \theta_0^{-1}$.  One has $s_{\rm diff} \propto \lambda^{-2/(\beta-2)}$ for $s_{\rm diff} > l_0$ and $s_{\rm diff} \propto \lambda^{-2}$ for $s_{\rm diff} < l_0$.
Centimeter-wavelength observations show that the apparent size of Sgr A* scales as $\lambda^{2.01 \pm 0.03}$ for $\lambda \geq 2\,$cm (e.g. Lo \et\ 1998, Bower \et\ 2004), which indicates either $\beta \approx 4$ or $ s z /L < l_0$.   However, in view of  the foregoing arguments, only the latter explanation is viable.  Scatter-broadening measurements at centimeter wavelengths yield no information on the spectral index of the electron density fluctuations.

We normalize each scattering model by finding the appropriate $C_N^2 \Delta L$ or $C_N^2(0)$ to reproduce the observed angular radius of the image of Sgr A*.  In the thin screen model the phase structure function at $r<l_0$ is approximated by 
\begin{eqnarray}
D_\phi ({\bf r}) &\approx& \frac{2^{5-\beta} \pi^{5/2} r_e^2 \lambda^2 l_0^{\beta-4} R^{\beta/2} {\rm SM} }{ \cos \left(\frac{\pi \beta}{2} \right) } \left\{ \frac{\Gamma \left(2-\frac{\beta}{2} \right)}{R^3 \, \Gamma \left( \frac{\beta}{2} \right) \Gamma \left(\frac{3-\beta}{2} \right) } \left[ 
\frac{r_y^2}{3-\beta}  \, \null_2 F_1 \left(\frac{3}{2},2-\frac{\beta}{2};\frac{5-\beta}{2};\frac{1}{R^2} \right) \right. \right. \nonumber \\
&\null& \left. \left. \qquad \qquad \qquad - R^2 r_x^2 \, \null_2 F_1 \left(\frac{1}{2},2-\frac{\beta}{2};\frac{3-\beta}{2};\frac{1}{R^2} \right)  \right]  \right. \nonumber \\
&\null& \left. + \frac{R^{-\beta}}{ \Gamma\left(\frac{\beta-1}{2} \right) } \left[ \frac{r_x^2}{\beta-1} \,  
\null_2 F_1 \left(\frac{3}{2},\frac{\beta}{2};\frac{1+\beta}{2};\frac{1}{R^2} \right) 
- r_y^2 \, \null_2 F_1 \left(\frac{1}{2},\frac{\beta}{2};\frac{\beta-1}{2};\frac{1}{R^2} \right)  \right]
\right\}  , r \ll l_0. \label{DphiAnisThin1}
\end{eqnarray}
% derived by expanding the $1-\exp (i {\bf q} \cdot {\bf r})$ part of the integrand  in eq. (\ref{DphiPrime}) to second order in ${\bf q}$.  When $r<l_0$ the $\exp[-(ql_0/2)^2]$ term cuts the integrand off before ${\bf q} \cdot {\bf r}$ reaches unity, yielding
This equation is solved in conjunction with eq. (\ref{MutCoher}) for a given $R$, $z$, $L$ and $\theta_0$ to find the scattering measure required to reproduce the angular size and ellipticity of the image of Sgr A*.   In the regime $r \ll l_0$ applicable to the scatter-broadening of Sgr A*, the value of $R$ and the image ellipticity do not correspond exactly, so one finds the value of $R$ numerically.
%Figure \ref{RvsZetaKolmog} shows the relationship between $R$ and $\zeta$.    

%In the two-screen model, with screens at distances $z_1$ and $z_2$ possessing scattering measures ${\rm SM}_1$ and ${\rm SM}_2$ respectively, the angular size of Sgr A* constrains only the combination 
%\begin{eqnarray}
%z_1^2 \, {\rm SM}_1  + z_2^2 \, {\rm SM}_2   = 
%\frac{2^{\alpha-1} L^2 \theta_0^2 l_0^{2-\alpha}}{\Gamma \left(1 - \frac{\alpha}{2} \right) r_e^2 \lambda^4}.
%\label{TwoScreenConstraint}
%\end{eqnarray}
In the extended medium model the integral $\int_0^L D({\bf s}_0z/L,z) dz$ can be inverted  when $\Delta z \la L$ and $s_0 \Delta z/L \la l_0$ to find $C_N^2(0)$ in closed form:
%\begin{eqnarray}
%\int_0^L D_\phi'({\bf s}z/L,z) dz = \frac{4 \pi^2}{2^\alpha} \Gamma \left(1 - \frac{\alpha}{2} \right) C_N^2 (0) r_e^2 \lambda^2 l_0^{\alpha-2} s^2 \frac{\Delta z^2}{L^2} \left[\sqrt{\pi} \Delta z {\rm erf}\left( \frac{L}{\Delta z} \right) - 2 L e^{-L^2/\Delta z^2} \right]
%\end{eqnarray}
\begin{eqnarray}
C_N^2(0) = \frac{2^{\alpha-1} \theta_0^2 L^2 l_0^{2- \alpha} }{\sqrt{\pi} r_e^2 \lambda^4 \Delta z^3 \Gamma(1-\alpha/2)}.
\end{eqnarray}
The models described below are normalized using the above expressions so that the scattering strength is always consistent with the observed degree of angular broadening toward Sgr A*.

Although angular broadening measurements leave the spectral index of turbulent electron density fluctuations unconstrained, we concentrate on only two specific cases that illustrate the range of behavior possible, $\beta=11/3$ and $\beta=3.9$.  The former case corresponds to Kolmogorov turbulence, while  the latter case is of interest because steeper spectra produce refractive modulations whose amplitude exhibits a weaker frequency dependence.  Such an index is suggested by the data, as the measured structure functions (Fig.\,\ref{StructureFns}) all saturate at similar amplitudes.  Qualitatively different refractive scintillation behavior is possible for yet steeper, $\beta \geq 4$ power spectra (e.g. Blandford \& Narayan 1985, Goodman \& Narayan 1985), but we do not consider such spectra here.  Scintillations in this regime cause a large degree of image wander which is incompatible with observed limits placed on this effect by VLBI astrometry on Sgr A* (Backer \& Sramek 1999; Reid \et\ 1999; Reid \& Brunthaler 2004).

% For R=1.
%If the visibilities decline on a baseline smaller than the inner scale of the medium (i.e. $s_{\rm obs} z/L < l_0$), then the observed angular broadening sets the constraint 
%\begin{eqnarray}
%C_N^2 \Delta L = \frac{2^{\alpha+1} \theta^2 L^2 }{\Gamma(1-\alpha/2) r_e^2 \lambda^4 l_0^{\alpha-2} z_0^2},
%\end{eqnarray}

\begin{figure}
\begin{center}
\includegraphics[scale=0.8]{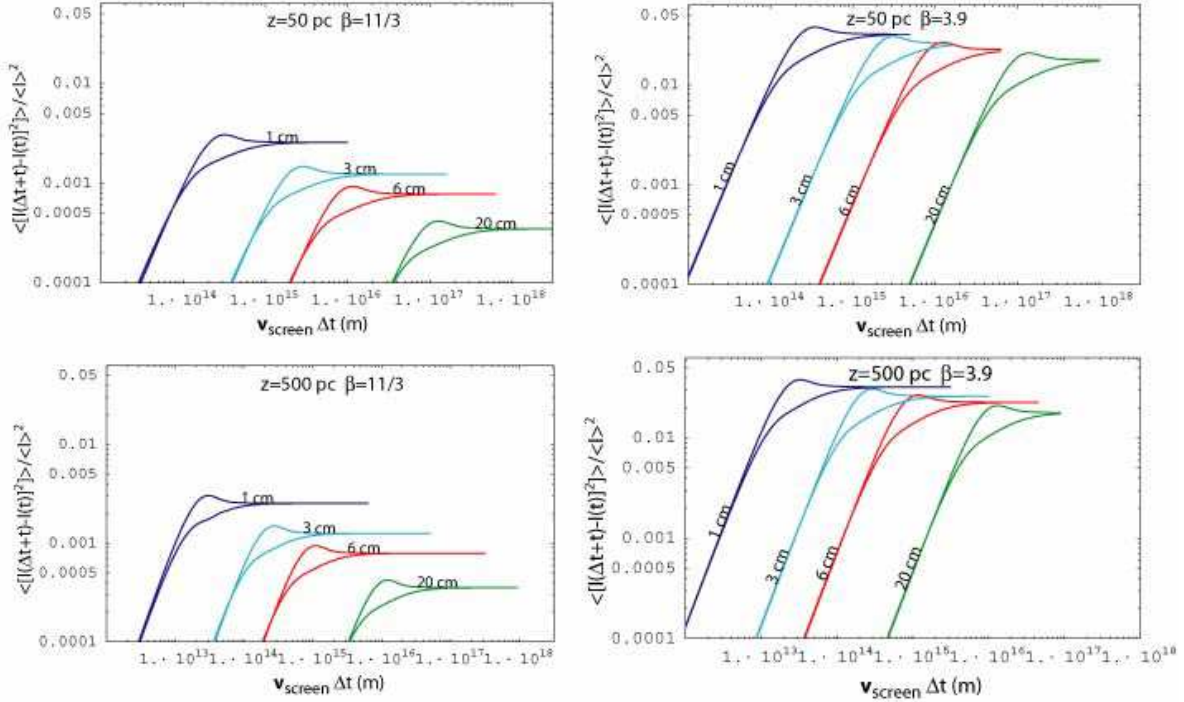}
\end{center}
% NB don't use DsThinxxx100 or 1000.jpg.eps because they were incorrectly normalized to Sgr A*'s
% scattering size.
\caption{Intensity structure functions from a thin-screen of scattering material with a turbulent power spectrum with index $\beta=11/3$ or $\beta=3.9$, located a distance of 50\,pc or 500\,pc from Sgr A*. The strength of the scattering is set by requiring the scattering to reproduce the observed degree of angular broadening of Sgr A*.  
%The colour coding represents the fluctuations at 1cm (blue), 3cm (light blue), 6cm (red) and 20cm (green).  
Because the turbulence is anisotropic the form of the intensity structure function depends on the orientation of the anisotropy axis with the scintillation velocity.  The two limiting cases are shown for each wavelength:  the curve which reaches the highest amplitude results when the velocity is oriented parallel to the major axis of the scattering disk, while the lower curve results when the velocity is oriented along the minor axis.  Equation (\ref{viss}) should be used to convert the values on the abscissae to units of time.} \label{ThinCIFig100}
\end{figure}

\subsection{Refractive Variations from a Thin Screen} \label{AnisThinScreen}

We quantify the amplitude and time scale of the intensity fluctuations expected from a thin scattering screen using the intensity autocovariance function, $C_I(t)$.  This quantity is directly related to the intensity structure function, $D_I(t) = 2 [C_I(0) - C_I(t)]$, so it permits direct comparison between the models and the observed variations.

Since the scattering medium is anisotropic the amplitude of the intensity variations on a given time scale depends on the relative orientation between the anisotropy and the scintillation velocity.  As the direction of the scintillation velocity is unknown, we calculate intensity structure functions both when the velocity is parallel and perpendicular to the anisotropy axis.  These two choices correspond to ${\bf v}_{\rm ISS} = (v,0)$ and ${\bf v}_{\rm ISS} = (0,v)$ respectively.
% since the power spectrum of electron density fluctuations was defined above such that the long axis of the source is oriented along the $x$-axis.  

The intensity autocovariance due to scattering from a thin screen is (Codona \et\ 1987, Coles \et\ 1987),
\begin{eqnarray}
C_I(\tau) &=& 8 \pi r_e^2 \lambda^2 {\rm SM} \int d^2{\bf q} \left(\frac{q_x^2}{R} + R q_y^2  \right)^{-\beta/2}
\left\vert V\left( \frac{{\bf q} (L-z) }{k} \right) \right\vert^2 \sin^2 \left[ \frac{q^2 z(L-z)}{2kL} \right] \nonumber \\ 
&\null& \qquad \qquad \qquad \qquad \times \exp \left[ -D_\phi \left( \frac{{\bf q} z(1-\frac{z}{L})}{k} \right) -  \left( \frac{q l_0}{2} \right)^2 + i  \frac{z \, \tau\, {\bf v}_{\rm ISS} \cdot {\bf q}}{L}   \right], \label{Cthin}
\end{eqnarray}
where the phase structure function for an anisotropic screen is given by eq. (\ref{DphiAnisThin1}) for $r \ll l_0$ and
\begin{eqnarray}
D_\phi ({\bf r}) =  - 2^{4-\beta} \pi^2 \, r_e^2 \, \lambda^2 \, {\rm SM} \,  \frac{ \Gamma \left(1- \beta/2 \right) }{ \Gamma \left( \beta/2 \right) } \, R^{1-\beta/2} (r_y^2 + R^2 r_x^2)^{\beta/2-1},\quad  2<\beta<4,
\end{eqnarray}
in the regime $r \ga l_0$.
% NB see Mma file AnistropicDphi.nb.  Note the minus sign out the front.  
% This is because Gamma[1-b/2]/Gamma[b/2] is negative.
The function $V({\bf r})$ is the source visibility.
We assume that the source is point-like and henceforth take $V=1$.  In practical terms, the visibility of the source is only important consideration {\it in the thin-screen model} if the intrinsic source angular diameter exceeds the angular broadening size.  The assumption of a point-like source is valid at centimeter wavelengths, as Lo \et\ (1998) and Bower \et\ (2004) report that intrinsic source size only becomes comparable to the scatter-broadening size at wavelengths shorter than 7\,mm.  

The scintillation velocity ${\bf v}_{\rm ISS}$ depends on the motions of the source and scattering material relative to the observer, ${\bf v}_{\rm src,o}$ and ${\bf v}_{\rm screen,o}$ respectively by (Gupta, Rickett \& Lyne 1994):
\begin{eqnarray}
{\bf v}_{\rm ISS} = \left\vert \left(1+\frac{L}{z}\right) {\bf v}_{\rm screen,o} - \left( \frac{L}{z} \right) {\bf v}_{\rm src,o} \right\vert. \label{viss}
\end{eqnarray}
Thus we see that the speed with which the scintillation pattern traverses the Earth is a factor $\sim L/z$ larger than the speed of the scattering material itself.  This correction factor is in the range $17-170$ for the plausible range of screen distances from Sgr A*.

Numerical integration is used to derive the structure function given the scattering measure, anisotropy ratio -- both derived from angular broadening measurements --  and a screen distance and spectral index, $\beta$.  The results for both $\beta=11/3$ and $\beta=3.9$ and $z=50\,$pc and $z=500\,$pc are shown in Figure \ref{ThinCIFig100}.  Table \ref{ModelsTable} lists the various models used and their associated parameters.

\begin{table}
\centerline{\title{Thin-screen model parameters}}
\begin{center}
%\begin{tabular}{|c|c|c|c|c|c|c|}
\begin{tabular}{|c|c|c|}
\tableline
%\null & \multicolumn{6}{c|}{$\beta$}  \\
\null & \multicolumn{2}{c|}{$\beta$} \\
\tableline
%screen distance (pc) & \multicolumn{3}{c|}{3.67} & \multicolumn{3}{c|}{3.9} \\
distance (pc) & 3.67 & 3.9 \\
\tableline
%\null & SM & sat. amp. & sat. $\tau$ & SM & sat. amp. & sat $\tau$ \\
\tableline
%50 & $2.70 \times 10^{25}$\,m$^{-5.67}$ &\null & \null  & $6.35 \times 10^{23}$\,m$^{-5.9}$ & \null & \null  \\
50 & SM$=2.70 \times 10^{25}$\,m$^{-5.67}$ & SM$=6.35 \times 10^{23}$\,m$^{-5.9}$ \\
%500 & $2.70 \times 10^{23}$\,m$^{-5.67}$ & \null& \null& $6.35 \times 10^{21}$\,m$^{-5.9}$ & \null & \null \\
500 & SM$=2.70 \times 10^{23}$\,m$^{-5.67}$ &  SM$=6.35 \times 10^{21}$\,m$^{-5.9}$  \\  
\tableline
\end{tabular}
\end{center}

\vskip 0.2 in

\centerline{\title{Extended medium model parameters}}
\begin{center}
%\begin{tabular}{|c|c|c|c|c|c|c|}
\begin{tabular}{|c|c|c|}
\tableline
%\null & \multicolumn{6}{c|}{$\beta$}  \\
\null & \multicolumn{2}{c|}{$\beta$} \\
\tableline
%screen thickness (pc) & \multicolumn{3}{c|}{3.67} & \multicolumn{3}{c|}{3.9} \\
screen thickness (pc) & 3.67 & 3.9 \\
\tableline
%\null & $C_N^2(0)$ & sat. amp. & sat. $\tau$ & $C_N^2(0)$ & sat. amp. & sat. $\tau$ \\
\tableline
%50 & $1.10 \times 10^7$\,m$^{-6.67}$ &\null & \null & $2.52 \times 10^{5}$\,m$^{-6.9}$ & \null & \null  \\
50 & $C_N^2(0)=1.10 \times 10^7$\,m$^{-6.67}$ & $C_N^2(0)=2.52 \times 10^{5}$\,m$^{-6.9}$  \\
%500 & $1.10 \times 10^4$\,m$^{-6.67}$ & \null & \null & $2.52 \times 10^{2}$\,m$^{-6.9}$ & \null & \null \\
500 & $C_N^2(0)=1.10 \times 10^4$\,m$^{-6.67}$ & $C_N^2(0)=2.52 \times 10^{2}$\,m$^{-6.9}$ \\
\tableline
\end{tabular}
\end{center}

\caption{The scattering parameters used for each model investigated. 
%The saturation amplitudes and time scales of the model lightcurve structure functions are listed (for a source size of $1\mu$as in the case of the extended medium models).  
All models assume a distance of $L=8.5\,$kpc to Sgr A* and an inner scale $l_0 =100\,$km.  In the thin screen model the scattering parameters are derived using the observed scatter-broadened diameter of Sgr A* at $\lambda 6.03\,$cm of 49.6\,mas along the major axis.  In the extended medium model, which does not take scattering anisotropy into account, the scattering parameters are matched to the average diameter, 37.5\,mas.} \label{ModelsTable}
\end{table}

\begin{figure}
\begin{center}
\includegraphics[scale=0.75]{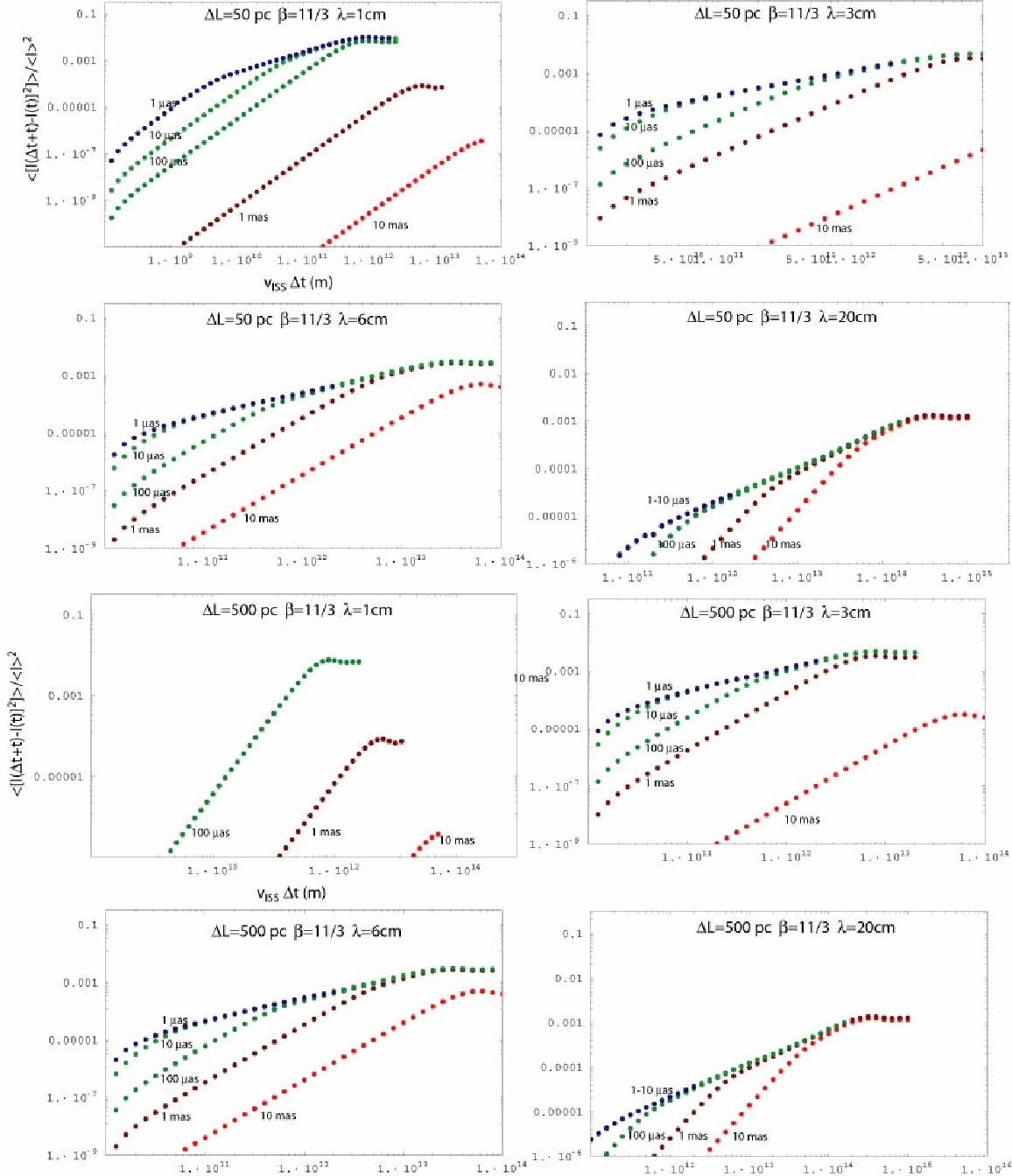}
\end{center}
\caption{Theoretical structure functions as a function of wavelength for the extended medium model for a Kolmogorov spectrum of turbulence ($\beta=11/3$) and for medium depths of $50$ and $500$\,pc.
The effect of finite size is illustrated by the various curves in each figure: red, brown, dark green, light green and blue dots represent the structure functions for a source of intrinsic angular size $10, 1, 0.1, 0.01$ and $0.001\,$mas respectively.  A finite turbulent dissipation scale of $10^5\,$m is assumed in the calculation. } \label{ExtCIFig1}
\end{figure}

\begin{figure}
\begin{center}
\includegraphics[scale=0.75]{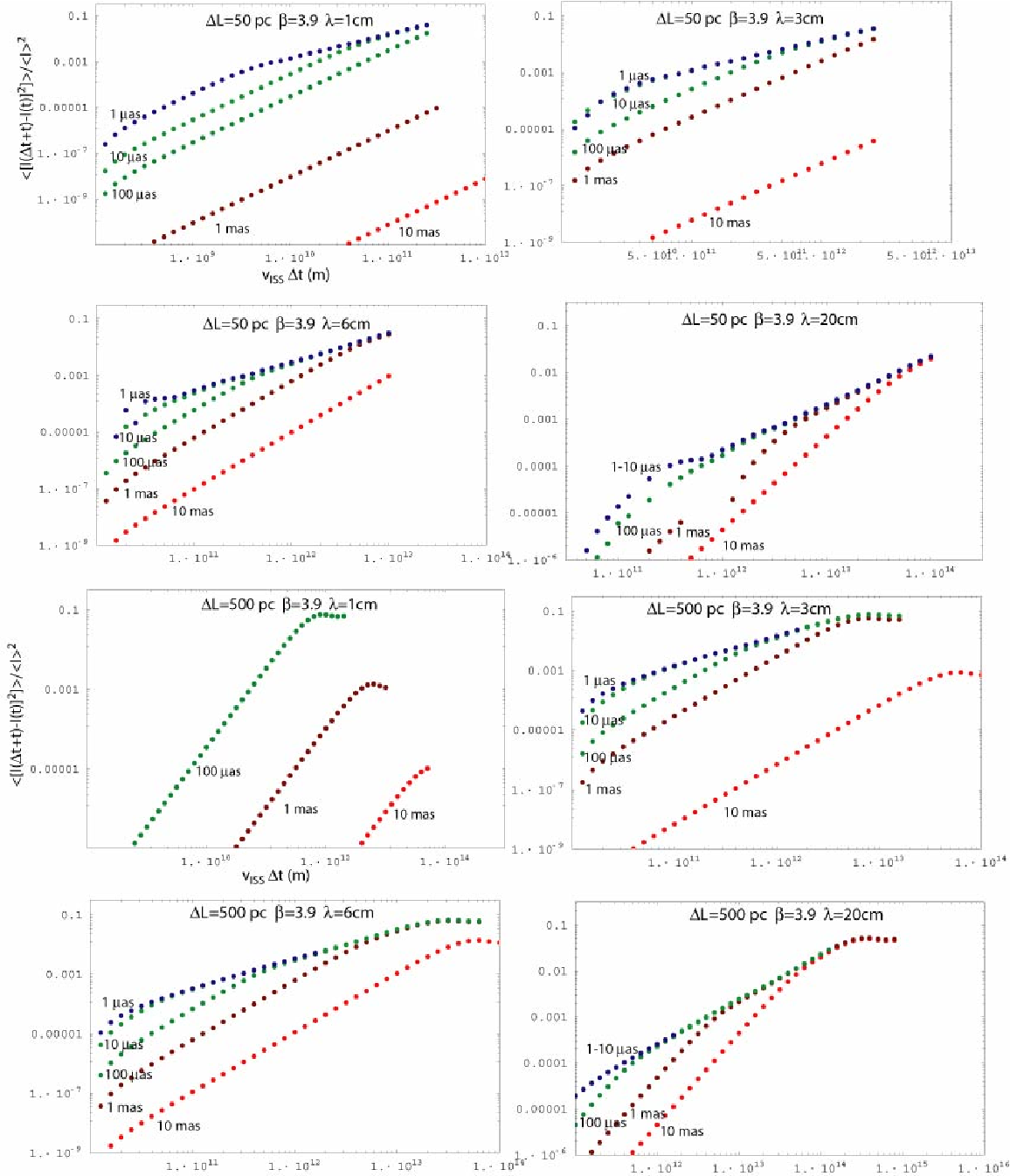}  
\end{center}
\caption{The same as Figure \ref{ExtCIFig1} but a steeper, $\beta=3.9$ electron density fluctuation power spectrum. } \label{ExtCIFig2}
\end{figure}

\begin{figure}
\centerline{\includegraphics[scale=0.9]{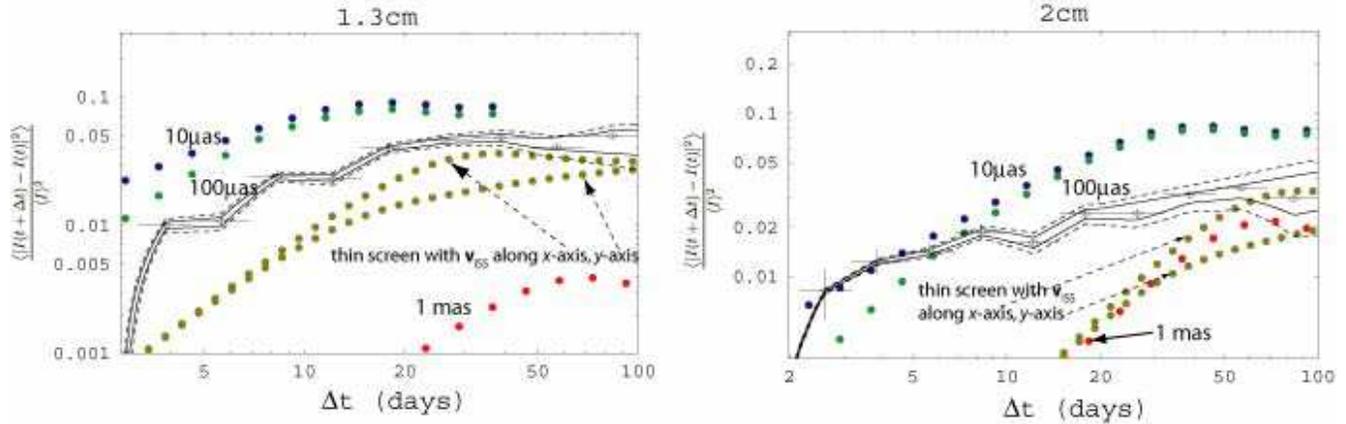} }
\caption{A comparison between the observed structure functions and those predicted by the extended-medium $\Delta =500\,$pc, $\beta=3.9$ model for a plausible range of source sizes, and the $\beta=3.9$, $z=500$ thin-screen model.  A scintillation speed $v_{\rm ISS}=1000\,$km\,s$^{-1}$ is assumed in this plot.} \label{DsComparison}
\end{figure}

\subsection{Refractive Variations from an Extended Medium} \label{ExtMedModel}
In an extended scattering medium the autocovariance of intensity fluctuations is given by (Codona \et\ 1987, Coles \et\ 1987)
\begin{eqnarray}
C_I(\tau) &=&  8 \pi r_e^2 \lambda^2 \int_0^L dz \int_{-\infty}^\infty d^2 {\bf q} \, 
\left\vert V\left( \frac{{\bf q} (L-z) }{k} \right) \right\vert^2 \, \Phi_{N_e}({\bf q},z) \, \sin^2 \left[ \frac{q^2 z (L-z)}{2 k L}\right] \nonumber \\
&\null& \qquad \qquad \qquad \times \exp\left[ -\int_0^L D_\phi' \left(\frac{q h(t,z)}{k} ,t \right)\,dt + i \frac{z {\bf q} \cdot {\bf v}_{\rm ISS} \tau }{L} \right]  \label{Crs}
\end{eqnarray}
where 
\begin{eqnarray}
h(t,z) = \left\{ \begin{array}{ll} 
t (z/L -1 ), & t<z, \\
z (t/L-1), & t >z. \\
\end{array} \right. 
\end{eqnarray}

The assumption that the source is point-like is somewhat more problematic when the scattering medium  extends close to the source, since clearly here the apparent angular source size must be appreciable relative to the local size of the scattering disk.  Moreover, it is possible for the source size to affect the amplitude of intensity fluctuations without exerting a strong influence on the shape of the scatter-broadened image.  This is because scattering material closer to the observer influences the angular broadening more strongly; eq. (\ref{MutCoher}) shows that one requires a smaller baseline ${\bf s}$ for a larger value of $z$ to satisfy $s z/L= s_{\rm diff}$.  

The effect of source size is investigated here by calculating the structure functions for a variety of intrinsic source sizes, from $1\,\mu$as to $10\,$mas.  The source is modeled as a circular disk whose visibility falls to zero on baselines longer than $1/k \theta_{\rm src}$ so that the term $V({\bf q} (L-z)/k)$ appearing in the integrand above is approximated as $H[ 1/(L-z) \theta_{\rm src} - q]$, where $H$ is the unit step function.
% i.e. H(x)=0 when x < 0 and =1 when x>1.  We want a function that is one only when q< 1/(L-z) \theta.
We also make the simplifying assumption that the scattering is always strong so that, even for very small $z$, the exponential in eq. (\ref{Crs}) cuts off the integrand before the sine-squared term begins to oscillate  (i.e. $r_{\rm diff}(z) < \sqrt{z/k}$).   We therefore expand the sine function for small argument and obtain the following expression for the intensity autocovariance from an extended scattering medium
\begin{eqnarray}
C_I (\tau) &=& \frac{r_e^2 \lambda^4 C_N^2(0) }{L^2} \int_0^L dz \, z^2 (L-z)^2 
e^{-z^2/\Delta z^2} \int_0^\infty dq \, q^{5-\beta} \, {\rm J}_0 \left( \frac{q \, v_{\rm ISS} \tau \, z}{L} \right) 
H \left[ \frac{1}{(L-z) \theta_{\rm src}} - q \right]  \nonumber \\
&\null& \qquad  \qquad \qquad \qquad \qquad  \times
 \exp\left[ - \left( \frac{q l_0}{2}\right)^2 -\int_0^L D_\phi' \left(\frac{q h(t,z)}{k} ,t \right)\,dt   \right]. \label{CIintermediate}
\end{eqnarray}
The integrand over $q$ cuts-off sharply once the source size becomes important, at $(L-z)^{-1} \theta_{\rm src}^{-1}$, or when either of the two arguments of the exponential exceed unity, at $2 l_0^{-1}$, or $q_\phi$ which is defined by the implicit equation 
\begin{eqnarray}
1 &=& \frac{8 \pi^2}{\beta-2} C_N^2(0) \Gamma \left(2 - \frac{\beta}{2} \right) r_e^2 \lambda^2 \left(\frac{l_0}{2} \right)^{\beta-2} \left\{
L \int_0^{z/L} e^{-t^2 {\cal L}^2} \null_1 F_1 \left[1-\frac{\beta}{2};1;- \left( \frac{q_\phi t (L-z)}{l_0 k} \right)^2 \right] dt \right. \nonumber \\ &\null& \left. \qquad \qquad  \qquad \qquad + 
L \int_{z/L}^1 e^{-t^2 {\cal L}^2} \null_1 F_1 \left[ 1-\frac{\beta}{2};1;- \left( \frac{q_\phi z (1-t)}{l_0 k} \right)^2 \right] dt
- \frac{\sqrt{\pi}\, {\rm erf}({\cal L})}{2} \right\},
\end{eqnarray}
where we write ${\cal L} = L/\Delta z$.  Noting that the scintillation velocity is well-approximated by  $|{\bf v}_{\rm srcreen,o} - {\bf v}_{\rm src,o}|L /z$ for the values of $L/z \gg 1$ under consideration here, the argument of the Bessel function in eq. (\ref{CIintermediate}) becomes $q v_{\rm eff} \Delta t$, where $v_{\rm eff}  = | {\bf v}_{\rm screen,o}-{\bf v}_{\rm src,o}|$.  The intensity autocovariance thus reduces to 
%\begin{mathletters}
\begin{eqnarray}
C_I(\tau) &=& \frac{r_e^2 \lambda^4 C_N^2(0)\, L^3}{6-\beta}  \int_0^1 dz \, z^2 (1-z)^2 
e^{-z^2 {\cal L}^2} q_{\rm max}^{6-\beta} \null_1 F_2 \left( 3-\frac{\beta}{2}; 1, 4- \frac{\beta}{2} ; - \frac{q_{\rm max}^2 v_{\rm eff}^2 \tau^2}{4}\right) ,  \nonumber \\
&\null&
\qquad \qquad \qquad  \qquad \qquad q_{\rm max} = \min[(L-z)^{-1} \theta_{\rm src}^{-1},2 l_0^{-1}, q_\phi], \quad  \beta < 6. \label{CrsIntegrate}
\end{eqnarray}
Below we also use eq. (\ref{CrsIntegrate}) to predict the amplitude of fluctuations at millimeter wavelengths.  Since the scattering is weaker at shorter wavelengths, an additional wavenumber cut-off, at the inverse of the Fresnel scale, $q_{\rm max} = \sqrt{4 \pi L/\lambda z (L-z)}$, is introduced so that the amplitude of the variability is not overestimated.  

%\begin{eqnarray}
%q_{\rm max} &=& 2 k \left[ \frac{\alpha\, \Gamma(1+\alpha/2) }{8 \pi^2 \Gamma(1-\alpha/2) C_N^2(0)
% r_e^2 \lambda^2 L^{1+\alpha} g(z) }\right]^{1/\alpha}, \\
%g(z) &=& \frac{1}{2} (1 -z)^\alpha {\cal L}^{-1-\alpha} 
%\left[ \Gamma \left( \frac{1+\alpha}{2}\right) - \Gamma \left(\frac{1+\alpha}{2},z^2 {\cal L}^2 \right)  \right] + z^\alpha \int_z^1  (1-t)^\alpha \, e^{-t^2 {\cal L}^2} \,dt . \label{Crsgz}
%\end{eqnarray} \end{mathletters}
% NB q_{\rm max} solution comes from assuming a negligible inner scale.  Which may or may 
% not be correct under these circumstances.
Equation (\ref{CrsIntegrate}) is integrated numerically to derive the intensity structure function expected due to refractive scintillation in an extended medium.  These functions are shown in Figs. \ref{ExtCIFig1} and \ref{ExtCIFig2}.

The time scale at which the structure functions saturate can be understood in terms of the time required for the turbulent medium to traverse the scattering disk, of order $L \theta_0/2 v_{\rm ISS}$.  A more rigorous estimate of the time scale is obtained by comparing the spatial wavenumber of the cut-off in eq. (\ref{Crs}) with the form of eq. (\ref{MutCoher}).  The power spectrum of intensity fluctuations cuts-off when the term $\int_0^L D_\phi' (q h(t,z)/k,t) dt$ reaches unity.  The smallest wavenumber cut-off occurs at the outer boundary of the scattering medium, when $z=\Delta L$ and $h(t,z) \approx t$.  Eq. (\ref{MutCoher}), which describes angular broadening due to the scattering medium, involves an integral of similar form.  By equating arguments in the two integrals one sees that the power spectrum cuts-off at $q=2/\theta_0 L$, where $\theta_0$ is the scatter-broadened size.  This corresponds to a time scale $\tau = L \theta_0/2 v_{\rm screen}$. 
% so one equates the arguments of the two integrals $1=\int_0^L D_\phi'(s z/L)dz = \int_0^L D_\phi'(q z/k)dz$ to show that $q=s_0 k/L$.

The intensity variations potentially yield information on the source size on scales well below the size of the scatter-broadened image.  For instance, observations at 1\,cm could distinguish between a source size of $1\,\mu$as and $10\,\mu$as simply on the basis of its variability properties.  This might seem surprising because such small angular diameters represent only 0.1 and 1\% respectively of the angular diameter of the scattering disk at this wavelength.  This sensitivity to detail well below the scatter-broadened size is possible because phase fluctuations along the path of propagation are weighted differently between observations of scatter-broadening, which measures a second-order moment of the wavefield, and intensity variability, which represents a fourth-order moment. 

The ability to distinguish between such small source sizes arises in any scattering medium that extends very close to the source.  Refractive modulations are strongest when the scattering strength is near unity.  The scattering strength can be expressed as the ratio of the Fresnel scale $r_{\rm F} = \sqrt{z (1-z/L)/k}$ to the transverse scale in the scattering medium over which the phase changes by one radian (i.e. the length scale, $r_{\rm diff}$, at which $ D_\phi(r_{\rm diff},z) =1$).   The scattering strength increases with distance from the source as both the Fresnel scale and the amount of scattering material encountered along the ray path increase.  Thus material close to the source can contribute greatly to the intensity variations provided that the source is sufficiently small that it does not substantially quench this contribution.

\section{The distinction between Scintillation-Induced and Intrinsic variations} \label{Comparison}
% \section{Il cimento dellÕ armonia e dellÕ inventione}
% The contest between Harmony and Invention
% Vivaldi op. 8.
% Having discussed a scintillation model, what definitely must be intrinsic to the source?
%--------------------------------------------------------------------------------------------

We now consider to what extent scintillation can account for the observed variability.  To qualify as a viable explanation of the variability on any given time scale it must account for a large fraction of the modulation amplitude at that time scale.  

By this criterion no thin-screen model constitutes a viable explanation of the intensity variations in Sgr A* on any time scale.  Larger variability amplitudes are predicted in this model when the scattering material is placed further from the source, but even when the screen is placed 500\,pc from Sgr A* the variability amplitude is more than a factor of two below that observed.  

%These models are thus also unable to account for any the intensity fluctuations on time scales shorter than the saturation time scale.  
Another shortcoming of the thin screen model is its failure to account for the slopes of the observed structure functions at small time lags.  None of the observed structure functions rise more steeply than $\tau^{1}$, whereas the models predict that they should rise as $\tau^{\beta-2}$.  Moreover, even if a thin-screen model were to reproduce the amplitudes and time scales at which the structure functions saturate it would still not explain any variations on shorter time scales.  There does not appear to be any range of time scales for which any of the structure functions rise as steeply as the thin-screen model predicts.  We thus conclude that thin-screen models constitute a poor explanation for the intensity variations observed on any time scale.

Extended medium models fare significantly better at explaining the amplitude of the observed intensity variations.  As can be seen from Fig.\,\ref{ExtCIFig1}, models with shallow electron density power spectra ($\beta \leq 11/3$) still fail to account for most of the intensity variations, particularly at long wavelengths.    However, the steep $\beta=3.9$ power spectrum model reproduces both the saturation amplitudes $\approx 0.1$ of the observed structure functions and their weak dependence on wavelength.   

These structure functions also saturate at roughly the time scale observed in the data.   
The saturation time scales in the models can be understood as the time required by the scattering medium to traverse the spatial extent of the scattering disk.  This rule of thumb estimates the time scale correct to within a factor of two relative to the time scales indicated by the structure functions in Figs. \ref{ExtCIFig1} and \ref{ExtCIFig2}.  
%Assuming the measured scatter-broadened size of 50\,mas at 6\,cm and that angular broadening measurements probe phase fluctuations within the inner scale of the turbulence, 
The predicted time scale is 
\begin{eqnarray}
\tau_0 \approx 
%\frac{L \theta_0}{2 v_{\rm screen}} = 
368 \left( \frac{L}{8.5\,{\rm kpc}} \right) 
\left( \frac{\lambda}{ 6\,{\rm cm}} \right)^2  
\left(\frac{v_{\rm screen}}{1000\,{\rm km\,s}^{-1}} \right)^{-1} \quad {\rm days}.
\end{eqnarray}

Certain extended-medium models also reproduce the generally shallow slope of the observed structure functions at small time lags.  This behavior depends on the source size relative to the angular scales probed by the scintillation.  Shallow slopes are present in some model structure functions over a large range of time scales, spanning up to two orders of magnitude in time scale before the structure function saturates. 

To illustrate the distinction between thin and extended-medium models, Fig.\,\ref{DsComparison} plots the observed 1.3 and 2\,cm structure functions against several models.  It is interesting to note that the $\beta=3.9$ extended-medium model with an intrinsic source size of $\sim 300\,\mu$as would appear to closely match the observed structure functions.  This size is comparable to that recently deduced by Bower \et\ (2004) using VLBI.

%Why does the structure function rise more slowly in certain portions?

Our conclusion is that extended medium models with steep $\beta \ga 3.9$ power spectra are capable of reproducing the gross features of the variability at centimeter wavelengths.  They do not explain {\it all} the detailed features of the structure functions.  In particular, they fail to account for any of the various peaks apparent in the structure functions in the range $50-100$\,days which, as discussed in \S\ref{VarCharacteristics}, which are probably due to flaring.    Although it is hard to see how these features could be reproduced by a simple scintillation model, it is nonetheless pertinent to consider how the assumptions used in the models bear on the predicted scintillation properties.  

In all extended medium models it is assumed that all layers of the scattering medium move with identical peculiar velocities.  When this is not the case different layers of the medium may cause intensity fluctuations on different time scales.  The extent to which this could occur depends on the relative contributions that layers at different distances make to the intensity fluctuations.  Significant variations will only be observed on different time scales if two layers which both contribute substantially to the scattering each possess different transverse velocities.  It is possible to see which layers contribute most to the scattering by comparing the amplitude of the structure functions for medium scale lengths of 50 and 500\,pc.  The intensity fluctuations are dominated by scattering layers within $50$\,pc from Sgr A* if the amplitude on a given time scale is identical in the two models.  Conversely, when they differ, most of the intensity fluctuations originate in the scattering medium beyond $50\,$pc from Sgr A*.  

The time scale may  also vary from the wavelength scaling predicted if different layers of the medium move at different velocities.  This is possible because the layer that contributes most to the intensity fluctuations changes with wavelength (see Fig.\,\ref{ExtCIFig2}). 

Another assumption inherent to the scattering model lies in the simplicity of the source structure.  Our models assume the simplest possible structure: a source comprised of one single circularly-symmetric component.  More complicated structure could cause qualitatively different variability.  To be specific, eqs. (\ref{Cthin}) and (\ref{Crs}) establish how the power spectrum of the intrinsic source brightness distribution, $|V({\bf r})|^2$ alters the power spectrum of scintillation-induced intensity fluctuations.  

Consider, for instance, how a source comprised of two compact components (e.g. a jet and a counterjet) would alter the scintillation characteristics.  Whereas the visibility of a single point source is constant and independent of baseline length,  the visibility amplitude of a double source with angular separation $\Delta \theta$ pointing along the direction $\hat{\btheta}=\Delta \btheta/\Delta \theta$ oscillates on a scale of length ${\bf r}_{\rm osc}=\hat{\btheta}/k \Delta \theta$.  This oscillation enhances the scintillation fluctuations at certain time scales relative to others.   The first peak of the oscillation emphasizes the power spectrum of intensity fluctuations at a fundamental spatial wavenumber ${\bf q}=\hat{\btheta} \, [\Delta \theta \, (L-z)]^{-1}$, for a scattering layer a distance $z$ from Sgr A*.  This corresponds to fluctuations on time scales $\tau_{0} \approx \Delta \theta (L-z)/v_{\rm screen}$.  The amplitudes of visibility oscillations at higher harmonics depend on the size of the components constituting the double source.  The visibility amplitude of a double source comprised of sufficiently compact components contains peaks comparable to the amplitude of the fundamental peak at integer multiples of ${\bf r}_{\rm osc}$.  
This in turn would enhance the scintillation fluctuations at shorter time scales, with the $n$th harmonic enhancing fluctuations on a time scale $\tau \approx \tau_{0}/n$ relative those on surrounding time scales.  
%This behavior is caused by oscillations in the visibility amplitude which boost scintillation power at multiples of the fundamental wavenumber.

It is possible in principle for source structure to explain additional features of the fluctuations observed toward Sgr A*.  One could appeal to a double source structure to explain peaks in the observed structure functions.  However, this model cannot not explain the peaks at time lags between 30 and 150\,days that are discussed in \S\ref{VarCharacteristics}.  A double source would give rise to a number of regularly-spaced peaks in the structure function, not a single, isolated peak.  Moreover, the source separation required to explain the location of the peaks is sufficiently large that it would have been observed.  A peak on a time scale of $\tau = 50\,$days would require a double source of separation $3\,$mas for a scattering medium moving at 1000\,km\,s$^{-1}$.  

We conclude that the interpretation of the isolated peaks present in the structure functions discussed in \S\ref{VarCharacteristics} as flares intrinsic to Sgr A* is robust to the assumptions made in the scintillation models considered here.

\subsection{The predicted role of scintillation at millimeter wavelengths}

Given the success of the $\beta=3.9$ extended medium model in explaining the broad characteristics of the centimeter variability, we have applied it to the predict the variations at millimeter wavelengths.  Sgr A* flux monitoring is already being carried out at the Sub-Millimeter Array (Zhao et al. 2003), and similar monitoring will soon be possible using CARMA and, eventually, ALMA.  

\begin{figure}
\centerline{\includegraphics[scale=0.8]{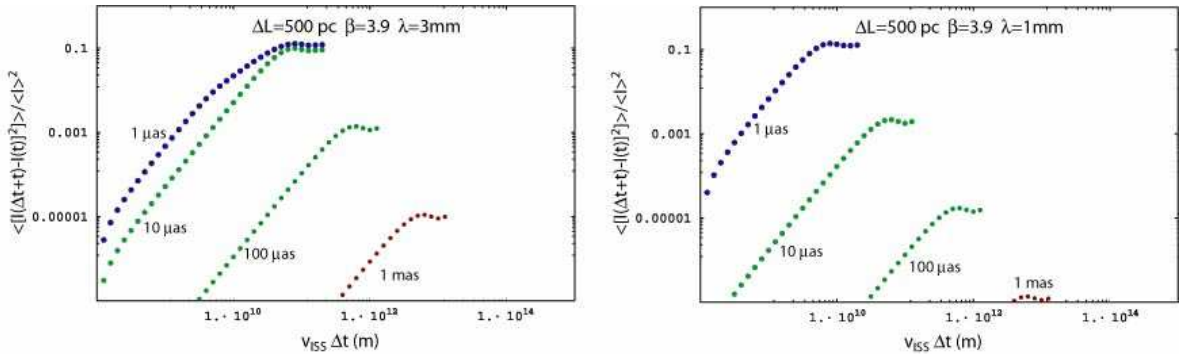}}
\caption{The structure function of the predicted variations for the best-fitting, $\Delta L=500\,$pc, $\beta=3.9$, scintillation model at centimeter wavelengths applied at 1 and 3\,mm.} \label{ExtCIPredict}
\end{figure}

The predictions at 1 and 3\,mm from the $\Delta L=500\,$pc, $\beta=3.9$ extended-medium model are shown in Fig. \ref{ExtCIPredict}.  This model implies that a $1\,\mu$as ($10\,\mu$as) source should exhibit 25\% (22\%) root-mean-square fluctuations on a time scale of $1.4 \, (v_{\rm screen}/1000\, {\rm km\,s}^{-1})$\,hours at 3mm.  A $100\,\mu$as would exhibit only 2.6\% variations, and on a time scale approximately three times longer.  Given recent measurements of the intrinsic size of Sgr A* at 7\,mm of $\approx 0.22$mas (Bower \et\ 2004) and assuming a $\nu^{-1}$ size-dependence it is reasonable to expect scintillation variations of order $3$\% at a wavelength of 3\,mm.
% NB Bower quotes a size of 24 Rs.  1 Rs for Sgr A* = 1.18 x 10^10m.  <=> which is 222.6 muas at
% 8.5 kpc.  For reference 

The scintillation characteristics are even more sensitive to source size at 1\,mm.  Only a source size of $\sim 1\,\mu$as is sufficiently small to exhibit 26\% fluctuations.  Sources of 10 and $100\,\mu$as would exhibit r.m.s.~fluctuations of 3.3 and 0.29\% respectively.  Again, the predicted variations occur on intra-day time scales of 4 and 12 hours respectively assuming a scintillation velocity of 1000\,km\,s$^{-1}$.

It is possible to already compare these predictions with a number of observations.  Both Wright \& Backer (1993) and Tsuboi \et\ (1999) report variations of order 1\,Jy amplitude at a wavelength of 3\,mm.   Such large flux density excursions are difficult to explain in terms of the present scintillation model, suggesting that intrinsic source activity is responsible for most of the variability.   

Mauerhan \et\ (2005) have recently claimed the Sgr A* undergoes only $\sim 20$\% intra-day variations at 3\,mm.  These variations, if real, would require an intrinsic source size of $10-30\,\mu$as at this wavelength to be consistent with scintillation.

%--------------------------------------------------------------------------------------------
%\section{Discussion} \label{Implications}
% DISCUSSION
%--------------------------------------------------------------------------------------------
%--------------------------------------------------------------------------------------------
\section{Discussion \& Conclusions} \label{Conclusions}
% CONCLUSIONS
%--------------------------------------------------------------------------------------------

Our analysis of multi-frequency monitoring data presented by Zhao \et\ (1992, 2001), Herrnstein 
\et\ (2004) and Falcke (1999) indicates that Sgr A* exhibits no quasi-periodic oscillatory behavior on any time scale between one week and 200 days.   The variability amplitudes are remarkably constant with frequency, varying between 30 and 39\%, but the time scale on which they saturate increases with wavelength.  

Several structure functions show evidence for variability on multiple time scales.  If the errors associated with the flux density measurements are correct, the data of Herrnstein \et\ (2004) indicate that the source exhibits unresolved 6-10\% inter-day ($<4$-day) variations between 7\,mm and 2\,cm.
No structure functions exhibit, within the errors, any evidence for appreciable variability with
time scales longer than 1000\,days.  The long term stability of the radio flux implies there is very little long-term variation in the accretion rate.   
% Short-term fluctuations can be driven by excitation of high energy electrons rather than accretion rate changes. 

We sought to explain the general features of the variability by reproducing the shape and amplitudes of the observed structure functions using several scintillation models.  Both thin-screen and extended-medium models were considered.  No thin screen model accounts for the properties of the variations.  The structure functions of the observed lightcurves rise less steeply with time lag and saturate at higher amplitudes than predicted.  They underpredict the amplitude of the variability by at least a factor of two at the saturation time scale and often by a more than an order of magnitude on shorter time scales.  If the medium responsible for the scattering of Sgr A* lies on a thin screen all of the observed flux variability must be intrinsic to the source itself.

Certain extended-medium models, on the other hand, do explain the amplitude of the fluctuations over a large range of time scales.  Models in which the electron density fluctuations follow a Kolmogorov power spectrum, corresponding to $\beta=11/3$, and a slightly steeper, $\beta=3.9$, power spectrum were investigated.  Of the two models, only those with a $\beta=3.9$ spectrum account for the amplitude of the fluctuations at all wavelengths.  If scintillation is to precisely predict the amplitude of the flux density variability of Sgr A*, this suggests that the power spectrum of the Galactic Center turbulence is slightly less steep, with an index lying in the range $3.8 \la \beta < 3.9$.

The most successful extended-medium model examined was used to predict the maximum contribution that scintillation could make to future observations of Sgr A* at millimeter wavelengths.  The expected variability amplitude depends strongly on the intrinsic source size.  A $1-10\,\mu$as object at 3\,mm would undergo fractional root-mean-square fluctuations of $\sim 25$\%, but a $100\,\mu$as source would exhibit only 3\% variations. Scintillation is even more sensitive to source size at 1\,mm, with a $1\,\mu$as source expected to display 26\% variations, but a $10\,\mu$as source would display only 3\% variability.

% [From Geoff: We should discuss or at least present physical conditions appropriate for our favored scintillating medium: density, or integrated column density.  Is there a known medium that we can associate this with?  The HIM?  What does it imply that we favor $\beta=3.9$?  

With what physical structures can we associate the extended scattering
medium?  Given only the approximate match between our model structure functions
and the actual structure functions, we do not expect that the extended
scattering medium must in fact be parameterized as discussed in the text.
In fact, the extended scattering medium might consist of only a few 
thin scattering media distributed over a range of distances from Sgr A*.
These thin media would have characteristics similar to the thin medium
discussed by Lazio \& Cordes (1998), with densities $>10^2 {\rm\ cm^{-3}}$.
The details of the structure functions are not sufficient to constrain
this result further.  For comparison purposes, we do compute the
mean density associated with the extended scattering medium as presented.
In this case, the mean density is  $\sim 10\,$cm$\,^{-3}$, assuming the outer scale of the turbulence to be 1\,pc.   This density is substantially greater than the density of the diffuse hot ionized gas ($T_e \sim 10^7\,$K, 
$n_e \sim 0.05 {\rm\ cm^{-3}}$) detected in the Galactic Center.  As Lazio \& Cordes (1998) discuss, 
the scattering medium may be the interface between this hot medium
and molecular clouds in the central 100 pc.  Our results are consistent with the picture reached by Lazio \& Cordes (1998) for the scattering medium with the modification that the scattering may take place at a range of distances from Sgr A*.

\acknowledgments

% \facility{HST(STIS)}, \facility{CXO(ASIS)}.



\clearpage

%% The following command ends your manuscript. LaTeX will ignore any text

\end{document}